\documentclass[preprint,12pt]{elsarticle}
\usepackage{lineno}

\usepackage{amsmath,amssymb,amsfonts,amsthm,algorithm }
\usepackage{float} 
\usepackage{graphicx}
\usepackage{booktabs}
\usepackage{siunitx}
\usepackage{xcolor}
\usepackage{pgfplots}
\usepackage[hidelinks]{hyperref}
\pgfplotsset{compat=1.18}
\usepackage{algpseudocode}
\usepackage{tabularx} 
\usepackage{comment}
\usetikzlibrary{calc}

\usepackage{listings}
\usepackage{xcolor}
\usepackage{subcaption}
\usepackage{longtable}
\usepackage{pdfpages}

\definecolor{codegray}{gray}{0.9}
\lstdefinestyle{mystyle}{
    backgroundcolor=\color{codegray},   
    basicstyle=\ttfamily\footnotesize,
    keywordstyle=\color{blue},
    commentstyle=\color{green!60!black},
    stringstyle=\color{red},
    numbers=left,
    numberstyle=\tiny\color{gray},
    stepnumber=1,
    breaklines=true,
    frame=single,
    captionpos=b,
    tabsize=4,
    showspaces=false,
    showstringspaces=false
}

\makeatletter
\def\ps@pprintTitle{%
  \let\@oddhead\@empty
  \let\@evenhead\@empty
  \def\@oddfoot{\reset@font\hfil\thepage\hfil}
  \let\@evenfoot\@oddfoot
}
\makeatother

\begin{document}

\begin{frontmatter}
\title{Parametrised-Quadratic Interface Reconstruction and Geometric Velocity Extension in the Volume-of-Fluid Framework}

\author[inst1,inst2]{V\,M\,S\,K Minnikanti}
\author[inst2]{Wenyi Yan}
\author[inst1,inst3]{Shyamprasad Karagadde\corref{cor1}}
\cortext[cor1]{Corresponding author}
\ead{s.karagadde@iitb.ac.in}

\address[inst1]{Department of Mechanical Engineering, Indian Institute of Technology Bombay, Mumbai, 400076 India}
\address[inst2]{Department of Mechanical and Aerospace Engineering, Monash University, 20 Exhibition Walk, Clayton VIC 3800, Australia}
\address[inst3]{Centre for Machine Intelligence and Data Science, Indian Institute of Technology Bombay, Mumbai, 400076 India}
\begin{abstract}

 A conservative Volume-of-Fluid (VOF) framework using local parametric-curve interface reconstruction (PI-VOF) is developed for geometric interface propagation on fixed Cartesian grids. While VOF methods achieve conservative and second-order accurate interface advection under prescribed velocity fields, normal-velocity-driven interface propagation requires the interfacial velocity to be extended to neighbouring staggered-grid locations.
 Existing studies on interface evolution employ auxiliary distance functions, kernel smoothers, or high-resolution VOF in multiple grids to compute propagation velocities and address first-order-accuracy limitations. In contrast, PI-VOF directly employs higher-order parametric interface geometry for local normal-velocity extension, aiming to achieve second-order accurate interface evolution on fixed Cartesian grids.
 The interface is represented by cell-wise parametric curves, with the prescribed volume fraction enforced through constrained optimisation. The reconstructed geometry is used to extend interface-normal velocities to neighbouring staggered-grid velocity locations through local scalar projection equations. The resulting velocity field is employed in conservative geometric advection using a directionally split Eulerian-implicit and Lagrangian-explicit (EI-LE) scheme, while the reconstructed interface provides the geometric information required for curvature evaluation and the ghost-fluid finite-difference discretisation of sharp interfacial Dirichlet conditions. Static reconstruction and reversible-vortex advection tests demonstrate higher-order interface geometry and approximately second-order transport convergence, with mass conservation to the chosen tolerance on refined grids. Prescribed normal-velocity and curvature-dependent propagation tests establish approximately second-order convergence of interface evolution. Coupled Stefan benchmarks involving capillarity and attachment kinetics demonstrate approximately second-order thermal convergence and decreasing interface errors under mesh refinement. Anisotropic dendritic solidification further demonstrates robust handling of grid-effects due orientation-dependent interface evolution, with tip velocities compared against microscopic solvability theory. The results establish a geometric formulation connecting conservative VOF transport, local normal-velocity extension, and sharp Stefan conditions for evolving interfaces.

\end{abstract}
\end{frontmatter}

\pagestyle{plain}




\section{Introduction}

Evolving interfaces arise in numerous multiphase flow and phase-change processes, including droplet coalescence, interfacial waves, atomisation, melting, solidification, and deposition~\cite{prosperetti2009multiphase,scardovelli1999interface}. Approaches to numerical studies of interface evolution fall broadly into two categories: \emph{interface tracking} and \emph{interface capturing}. Interface-tracking methods provide an explicit representation of the boundary and can achieve high accuracy for smooth interfaces~\cite{schmidt1994,quinghai-4th-track-2014}, but require additional treatment for topological changes. In contrast, interface-capturing approaches, such as the Volume-of-Fluid (VOF)~\cite{hirt1981volume,tryggvason2011direct} and Level Set (LS)~\cite{osher1988fronts,sussman1994level,level-set-review-gibou-2018} methods, represent interfaces implicitly through discrete fields, allowing changes in interface topology without explicit connectivity tracking and the imposition of boundary conditions in embedded-domain solvers~\cite{fedkiw1999non,li2006immersed,ghost2002}. In particular, for Stefan-type solidification problems, the interface velocity follows from the jump in thermal fluxes, while curvature- and normal-dependent effects such as surface tension and attachment kinetics can further influence the interfacial temperature. The LS method~\cite{Kim20000} is particularly suited to this task by constructing the extended velocity geometrically or through auxiliary extension equations or fast-marching, while it may demand reinitialisation and sometimes mass-correction.

The VOF method offers an alternative interface-capturing framework in which phase volumes are transported conservatively. In geometric VOF methods, interface reconstruction determines the location and orientation of the phase boundary from cell-wise volume fractions, providing the geometry required for flux computation and interface propagation. For geometric VOF methods, constructing a consistent extension directly from locally reconstructed interfaces is less straightforward, particularly when the interface orientation varies within a cell. Conventional reconstruction commonly employs piecewise-linear interfaces~\cite{Pilliod2004}, while higher-order reconstruction approaches have also been developed for smooth geometries~\cite{Scardovelli2003,Lopez2024Isosurface}. Moment-of-Fluid (MOF) methods provide further geometric information through additional moment data~\cite{SHASHKOV-2023-mof}. Although piecewise-linear reconstruction is effective for conservative transport, its constant within-cell orientation does not directly represent the variation of interface normals along a curved boundary. This becomes relevant when the propagation velocity or interfacial boundary conditions depend on local curvature and orientation. Such smoother normals for extension in VOF were achieved using reconstructed distance-function-based velocity evaluation, kernel-based smoothing, or volume-fraction-gradient-based propagation, beyond the reconstructed interface~\cite{lopez2010vof_JCP_solid,Lopez2013TwoGrid,Reitzle2017,Fakhreddine2024}, with the resulting interface propagation often limited to first-order accuracy. A higher-order geometric representation offers an alternative to piecewise-linear interfaces on fixed grids to improve the accuracy of VOF interface evolution from first to second order, without requiring a separate, highly resolved interface grid.

\begin{figure}[H]
    \centering
    \begin{tikzpicture}
        \node[anchor=south west, inner sep=0] (img)
        {\includegraphics[trim={4cm 0cm 4cm 0cm},width=10cm]{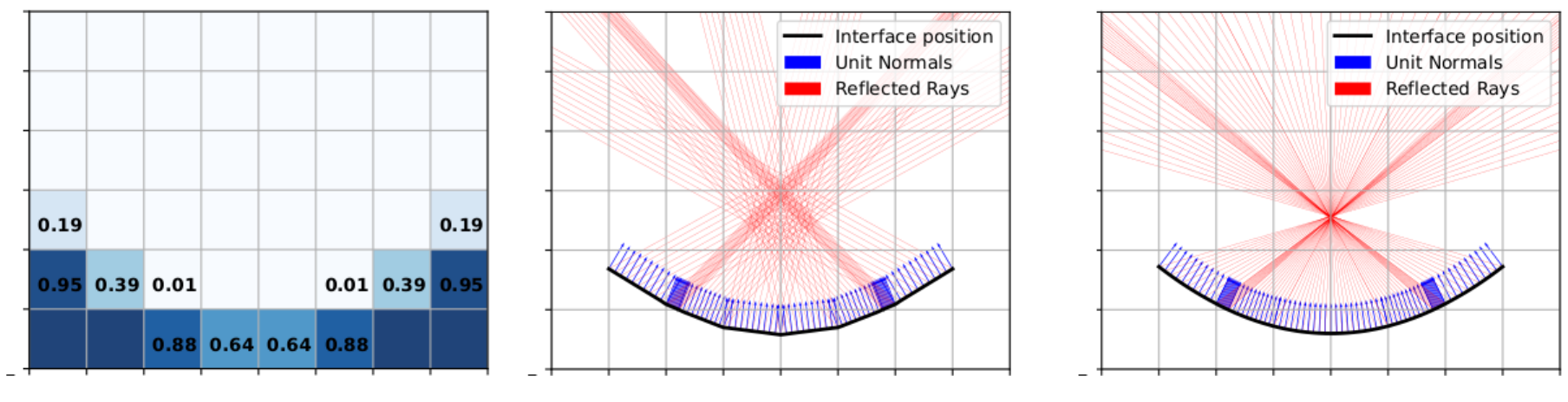}};
        
        \begin{scope}[x={(img.south east)}, y={(img.north west)}]
            \node at (0.05,-0.2) {(a)};
            \node at (0.50,-0.2) {(b)};
            \node at (0.92,-0.2) {(c)};
        \end{scope}
    \end{tikzpicture}
    \vspace{4pt}

    \caption{
    Illustrative comparison of local interface representations:
    (a) volume-fraction field containing a curved interface;
    (b) piecewise-linear reconstruction with constant within-cell interface orientation;
    (c) piecewise-quadratic reconstruction retaining the variation of interface orientation along the reconstructed curve. Incident and reflected rays are shown schematically to illustrate differences in the local interface orientation.
    }
    \label{fig:tower}
\end{figure}
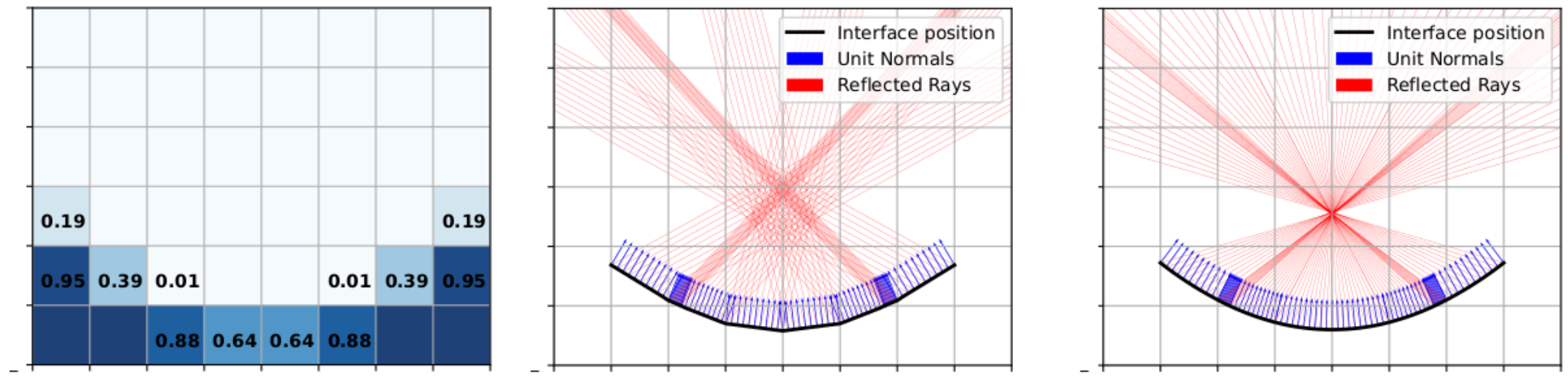

In this work, we present a non-linear Parametrised-quadratic Interface-reconstruction for the Volume-of-Fluid method (PI-VOF), coupled with a geometric normal-velocity extension procedure on fixed Cartesian grids. The reconstruction is formulated as a local constrained optimisation of the quadratic interface geometry, while narrow-band velocity extension is obtained separately through direct geometric projection onto the reconstructed segments. This enables the interface-normal velocities determined by Stefan-type conditions to be transferred to Cartesian velocity locations without solving an auxiliary extension PDE. As illustrated in Fig.~\ref{fig:tower} (focus of parabola is blurred by linear interfaces compared to PI-VOF), the quadratic representation provides geometric information along the reconstructed curve that is not available from a single linear segment. 

 The parametric representation is further used for local curvature and to extend interface-normal velocities into a narrow band. The extended velocity field is employed in conservative geometric VOF advection, and a ghost-fluid finite-difference discretisation is coupled to the reconstruction for imposing sharp Dirichlet and Stefan-type boundary conditions. The accuracy of the formulation is evaluated progressively through static reconstruction, reversible-vortex advection, prescribed normal-velocity propagation, and curvature-dependent interface evolution. Its coupling with thermal diffusion and interfacial balance conditions is examined using Stefan benchmarks with and without capillarity and attachment kinetics. Finally, anisotropic dendritic solidification is used to assess the sensitivity of the geometric velocity extension to interface orientation relative to the Cartesian grid, with dendrite-tip velocities compared against predictions from microscopic solvability theory. The present study is restricted to two dimensions and relies on linear-interface reconstruction using an isocontour-based approach~\cite{MinnikantiEICRUnderReview,Karagadde2024} to initialise the quadratic reconstruction and to handle limiting configurations, which can handle multiple material interfaces; however, topological changes are not considered in this study, and the analysis is limited to sufficiently smooth interfaces.

\section{Methods}

\subsection{Parametric quadratic reconstruction with geometric regularisation}
\label{subsec:reconstruction}

Let $\alpha_{ij}\in[0,1]$ denote the volume fraction of the reference phase in a Cartesian control volume $\Omega_{ij}$. Interface reconstruction is performed in mixed cells satisfying
\begin{equation}
    \varepsilon_{\alpha}<\alpha_{ij}<1-\varepsilon_{\alpha},
    \qquad \varepsilon_{\alpha}=10^{-8}.
    \label{eq:mixed_cell_definition}
\end{equation}
The reconstruction is performed in cell-local coordinates, with $\Omega_{ij}=[-1/2,1/2]^2$, and the resulting geometry is subsequently transformed to physical coordinates.

\begin{figure}[H]
\centering
\begin{minipage}[t]{0.38\linewidth}
\centering
\begin{tikzpicture}[scale=1.05]
    \draw[->,thick] (-0.15,0) -- (3.65,0);
    \fill (0,0) circle (1.5pt);
    \fill (3.4,0) circle (1.5pt);
    \node[below] at (0,0) {$\zeta=0$};
    \node[below] at (3.4,0) {$\zeta=1$};
    \draw[->,thick] (1.7,0.55) -- (1.7,1.35);
    \node[above] at (1.7,1.35) {$\mathbf X(\zeta)$};
\end{tikzpicture}

\vspace{2pt}
{\small (a)}
\end{minipage}
\hfill
\begin{minipage}[t]{0.58\linewidth}
\centering
\begin{tikzpicture}[scale=1.05]
    \draw[thick] (0,0) rectangle (4,3);

    \coordinate (Pzero) at (0,0.65);
    \coordinate (Pone)  at (4,2.20);
    \draw[very thick,blue]
        (Pzero) .. controls (1.20,1.95) and (2.75,2.55) .. (Pone);
    \fill (Pzero) circle (2pt);
    \fill (Pone) circle (2pt);
    \node[left] at (Pzero) {$\mathbf p_0$};
    \node[right] at (Pone) {$\mathbf p_1$};

    \draw[->,thick,blue] (Pzero) -- ++(0.90,0.88);
    \node[below right] at ($(Pzero)+(0.82,0.76)$) {$\mathbf t_0$};
    \draw[->,thick,red] (Pzero) -- ++(-0.62,0.64);
    \node[left] at ($(Pzero)+(-0.57,0.62)$) {$\mathbf n_0$};

    \draw[->,thick,blue] (Pone) -- ++(0.82,-0.28);
    \node[below] at ($(Pone)+(0.66,-0.25)$) {$\mathbf t_1$};
    \draw[->,thick,red] (Pone) -- ++(0.30,0.88);
    \node[right] at ($(Pone)+(0.25,0.78)$) {$\mathbf n_1$};
\end{tikzpicture}

\vspace{2pt}
{\small (b)}
\end{minipage}
\par\medskip

\begin{minipage}[t]{0.48\linewidth}
\centering
\begin{tikzpicture}[scale=1.00]
    \draw[thick] (0,0) rectangle (4,3);

    \coordinate (Pzero) at (0,0.75);
    \coordinate (Pone)  at (4,2.15);

    \draw[very thick,orange] (0,0.20) -- (0,1.35);
    \draw[very thick,orange] (4,1.45) -- (4,2.80);

    \draw[<->,thick] (0,0.20) -- (Pzero)
        node[midway,left] {$\delta_0^{-}$};
    \draw[<->,thick] (Pzero) -- (0,1.35)
        node[midway,left] {$\delta_0^{+}$};

    \draw[<->,thick] (4,1.45) -- (Pone)
        node[midway,right] {$\delta_1^{-}$};
    \draw[<->,thick] (Pone) -- (4,2.80)
        node[midway,right] {$\delta_1^{+}$};

    \draw[very thick,blue]
        (Pzero) .. controls (1.0,1.95) and (2.75,2.50) .. (Pone);

    \fill (Pzero) circle (2pt);
    \fill (Pone) circle (2pt);
    \node[right] at ($(Pzero)+(0.08,0)$) {$\mathbf p_0$};
    \node[left]  at ($(Pone)+(-0.08,0)$) {$\mathbf p_1$};
\end{tikzpicture}

\vspace{2pt}
{\small (c)}
\end{minipage}
\hfill
\begin{minipage}[t]{0.48\linewidth}
\centering
\begin{tikzpicture}[scale=1.00]
    \draw[thick] (0,0) rectangle (4,3);

    \coordinate (Pzero) at (0,0.75);
    \coordinate (Pone)  at (3.10,3);

    \draw[very thick,orange] (0,0.20) -- (0,1.35);
    \draw[very thick,orange] (2.35,3) -- (3.75,3);

    \draw[<->,thick] (0,0.20) -- (Pzero)
        node[midway,left] {$\delta_0^{-}$};
    \draw[<->,thick] (Pzero) -- (0,1.35)
        node[midway,left] {$\delta_0^{+}$};

    \draw[<->,thick] (2.35,3) -- (Pone)
        node[midway,above] {$\delta_1^{-}$};
    \draw[<->,thick] (Pone) -- (3.75,3)
        node[midway,above] {$\delta_1^{+}$};

    \draw[very thick,blue]
        (Pzero) .. controls (1.25,2.15)  .. (Pone);

    \fill (Pzero) circle (2pt);
    \fill (Pone) circle (2pt);
    \node[right] at ($(Pzero)+(0.08,0)$) {$\mathbf p_0$};
    \node[below] at ($(Pone)+(0,-0.08)$) {$\mathbf p_1$};
\end{tikzpicture}

\vspace{2pt}
{\small (d)}
\end{minipage}
\caption{Quadratic parametric reconstruction: (a) parameter interval $\zeta\in[0,1]$; (b) corresponding interface segment in cell-local coordinates.(c) admissible endpoint displacements on opposite cell faces; and
(d) admissible endpoint displacements on adjacent cell faces. The orange
segments denote the bounded intervals over which the endpoints may move}
\label{fig:parametric_space}
\end{figure}
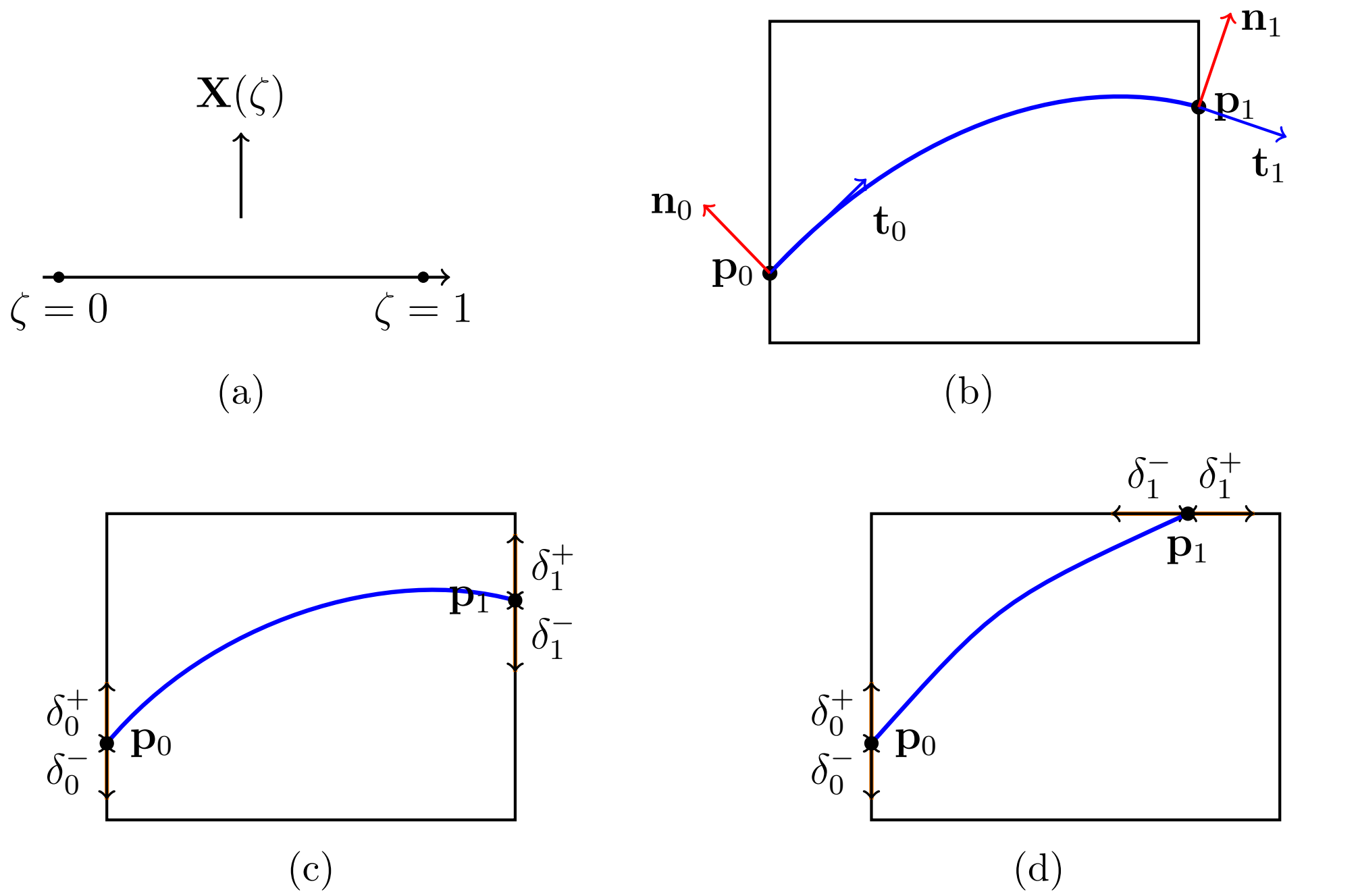

We adopt the Nagata-type quadratic parametric interpolation~\cite{Nagata2005}, developed for geometric modelling using endpoint positions and normals, and apply it to reconstructing interfaces from VOF data. A piecewise-linear isocontour identifies the intersected cell faces and supplies the initial interface intercepts~\cite{MinnikantiEICRUnderReview,Karagadde2024}. Let $\mathbf p_0$ and $\mathbf p_1$ denote the adjusted intercepts, with corresponding phase-oriented unit normals $\mathbf n_0$ and $\mathbf n_1$. The quadratic interface is represented as
\begin{equation}
    \mathbf X(\zeta)
    =\mathbf p_0+\mathbf A\zeta
    +\frac{\mathbf B-\mathbf A}{2}\zeta^2,
    \qquad 0\leq\zeta\leq1,
    \label{eq:quad_param}
\end{equation}
where
\begin{equation}
    \mathbf A=\lambda_0\mathbf t_0,
    \qquad
    \mathbf B=\lambda_1\mathbf t_1,
    \qquad
    \mathbf t_k=(-n_{k,y},n_{k,x}).
    \label{eq:quad_tangents}
\end{equation}
The tangent coefficients $\lambda_0$ and $\lambda_1$ are determined by enforcing $\mathbf X(1)=\mathbf p_1$, giving
\begin{equation}
    \begin{bmatrix}\mathbf t_0&\mathbf t_1\end{bmatrix}
    \begin{bmatrix}\lambda_0\\[1mm]\lambda_1\end{bmatrix}
    =2(\mathbf p_1-\mathbf p_0).
    \label{eq:quad_endpoint_constraint}
\end{equation}
Thus, $\mathbf X'(0)=\mathbf A$ and $\mathbf X'(1)=\mathbf B$ define the endpoint tangents. The parametric construction is illustrated in Fig.~\ref{fig:parametric_space}.

The endpoint normals are computed using compact height functions constructed from neighbouring volume fractions. The dominant coordinate direction is selected from the local volume-fraction variation, and interface heights are assembled from adjacent rows or columns. Centred finite differences of the heights provide the interface slopes, from which the phase-oriented unit normals are evaluated as
\begin{equation}
    \mathbf n
    =\frac{(-h',1)}{\sqrt{1+(h')^2}}
    \qquad\text{or}\qquad
    \mathbf n
    =\frac{(1,-h')}{\sqrt{1+(h')^2}}.
    \label{eq:recon_normals}
\end{equation}
A neighbour-consistency correction transfers compatible endpoint position and normal information between adjacent mixed cells through two local sweeps.

The two endpoint positions are constrained to their respective intersected cell faces. Each endpoint is expressed as
\begin{equation}
    \mathbf p_k(\delta_k)
    =\mathbf p_k^{\mathrm{lin}}+\delta_k\mathbf e_k,
    \qquad
    \delta_k^{-}\leq\delta_k\leq\delta_k^{+},
    \label{eq:face_intercept_parameter}
\end{equation}
where $\mathbf p_k^{\mathrm{lin}}$ is the initial guess's intersection, $\mathbf e_k$ is the unit vector along the corresponding face, and $\delta_k$ is the intercept correction that limits the search to the corresponding edge. The quadratic reconstruction is therefore determined by two unknowns, $(\delta_0,\delta_1)$ on two edges of the given cell.

For each trial pair of intercepts, the quadratic coefficients are obtained from Eq.~\eqref{eq:quad_endpoint_constraint}. The reconstructed phase area is calculated using the oriented boundary integral
\begin{equation}
    \widehat{\alpha}(\delta_0,\delta_1)
    =\frac{1}{2|\Omega_{ij}|}
    \oint_{\partial\Omega_{ij}^{\alpha}}
    \left(x\,dy-y\,dx\right),
    \label{eq:reconstructed_fraction}
\end{equation}
where $\partial\Omega_{ij}^{\alpha}$ consists of the quadratic interface and the cell-edge segments enclosing the reference phase. The parameter interval is partitioned at the required geometric intersections and turning points, and the contributions from the quadratic and cell-edge segments are integrated and summed to obtain the reconstructed volume fraction.
\paragraph{Mass-constrained selection of the quadratic segment}
The two endpoint corrections $(\delta_0,\delta_1)$ are unknown for the given volume fraction $\alpha$. Since the values can lie on the edge of the cell, the bounded intercept search can contain multiple mass-conserving solutions, including degenerate quadratics that collapse onto an endpoint-tangent line and curves that do not conform to the orientation of the endpoint normals. A geometric regularisation functional is therefore introduced to handle the degenerate and coherently oriented limits.
The intercepts are determined by imposing the prescribed volume fraction together with a geometric normal-tangent consistency objective. Defining the parametric tangent as,
\begin{equation}
    \mathbf q(\zeta)=\mathbf X'(\zeta)
    =\mathbf A+(\mathbf B-\mathbf A)\zeta,
    \label{eq:parametric_tangent}
\end{equation}
and the dimensionless regularisation functional is
\begin{equation}
    \mathcal R(\delta_0,\delta_1)
    =\int_0^1
    \frac{
    \sqrt{\left(\mathbf n_0\cdot\mathbf q\right)^2
         +\left(\mathbf n_1\cdot\mathbf q\right)^2}}
    {\|\mathbf q\|+\delta_n}\,d\zeta,
    \qquad \delta_n=10^{-15}.
    \label{eq:geometric_regularisation}
\end{equation}
The functional (Eq. \eqref{eq:geometric_regularisation}) measures the deviation of the parametric tangent from the endpoint-normal constraints along the reconstructed segment. The value $\delta_n$ is added to prevent division by zero. For parallel endpoint normals, its value is zero and falls back to the initial linear guess. 

For non-parallel endpoint normals, the regularisation suppresses collapse of the reconstruction onto either endpoint-tangent line. Neglecting only the division guard $\delta_n$, the square of the integrand in Eq.~\eqref{eq:geometric_regularisation} can be written as
\begin{equation}
\left[
\frac{\sqrt{(\mathbf n_0\cdot\mathbf t)^2+(\mathbf n_1\cdot\mathbf t)^2}}
{\|\mathbf q\|}
\right]^2
=
\frac{[1-(\mathbf n_0\cdot\mathbf n_1)^2][(1-\zeta)^2\lambda_0^2+\zeta^2\lambda_1^2]}
{(1-\zeta)^2\lambda_0^2+\zeta^2\lambda_1^2
+2\zeta(1-\zeta)\lambda_0\lambda_1
(\mathbf n_0\cdot\mathbf n_1)}.
\label{eq:regularisation_integrand}
\end{equation}
If $\lambda_0=0$ or $\lambda_1=0$, the quadratic degenerates into the line associated with $\mathbf t_1$ or $\mathbf t_0$, respectively, and the fraction in Eq.~\eqref{eq:regularisation_integrand} equals unity throughout the segment interior. For a coherently oriented quadratic, $\lambda_0\lambda_1>0$, the additional term in the denominator is less than one. 
Hence, minimising $\mathcal R$ therefore suppresses such degenerate configurations that don't conform to orientation near the endpoints
and selects the volume-conserving parametric segment compatible with the two
endpoint normals.

The volume-fraction equality constraint is scaled as
\begin{equation}
    \mathcal C(\delta_0,\delta_1)
    =c_V\left[
      \widehat{\alpha}(\delta_0,\delta_1)-\alpha_{ij}
    \right]    
    \label{eq:scaled_volume_constraint}
\end{equation}
The scaling factor $c_V$ adjusts the numerical magnitude of the volume residual supplied to the optimiser for the chosen tolerance (in the present work, uniformly $10^{-8}$ is used). The reconstruction is obtained by solving
\begin{equation}
    \begin{aligned}
    &\underset{\delta_0,\delta_1}{\operatorname{minimise}}
    &&\mathcal R(\delta_0,\delta_1),\\
    &\text{subject to}
    &&\mathcal C(\delta_0,\delta_1)=0,\\
    &&&\delta_k^{-}\leq\delta_k\leq\delta_k^{+},
    \qquad k=0,1.
    \end{aligned}
    \label{eq:recon_optimisation}
\end{equation}
The constrained optimisation is solved independently in each mixed cell using sequential least-squares quadratic programming (SLSQP)~\cite{Kraft1988SLSQP}. The linear-isocontour reconstruction~\cite{MinnikantiEICRUnderReview} is retained for configurations as a fallback to the quadratic solver.

\subsection{Conservative Advection: Eulerian implicit - Lagrangian explicit}
\label{subsec:advection}

The parametrised-quadratic interfaces reconstructed from the cell-centred volume fractions are transported using the Eulerian implicit--Lagrangian explicit (EI--LE) geometric advection scheme~\cite{Scardovelli2003,Ye2023}. The scheme applies successive directional remappings to the reconstructed phase regions, with their overlap areas determining the updated volume fractions. Here, $u_1,u_2$ and $v_1,v_2$ denote the face-normal displacements in the $x$- and $y$-directions, respectively, while $(x',y')$, $(x,y)$ and $(x'',y'')$ denote the coordinates before, between and after the directional remappings.
 Starting with an Eulerian implicit (EI) step in the $x$-direction, the departure region $\Omega_{\mathrm{depart}}$ is mapped into the intermediate region $\Omega_{i,j}$ as
\begin{equation}
\Pi_{x} : \Omega_{\mathrm{depart}} \to \Omega_{i,j}, \qquad
\begin{cases}
x = \beta_x(x' + u_1), \\
y = y',
\end{cases}
\label{eq:EI_step}
\end{equation}
where $\beta_x = \tfrac{1}{1+u_1-u_2}$.  

\begin{figure}[H]
  \centering
  \begin{minipage}[t]{0.49\linewidth}
    \centering
    \includegraphics[width=\linewidth]{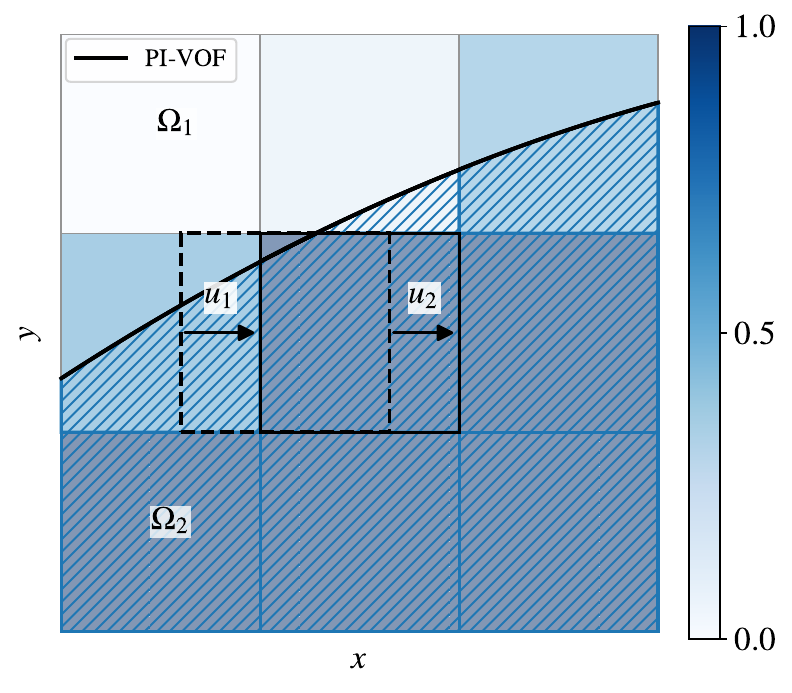}
    \par\smallskip\centering\small (a)
  \end{minipage}\hfill
  \begin{minipage}[t]{0.49\linewidth}
    \centering
    \includegraphics[width=\linewidth]{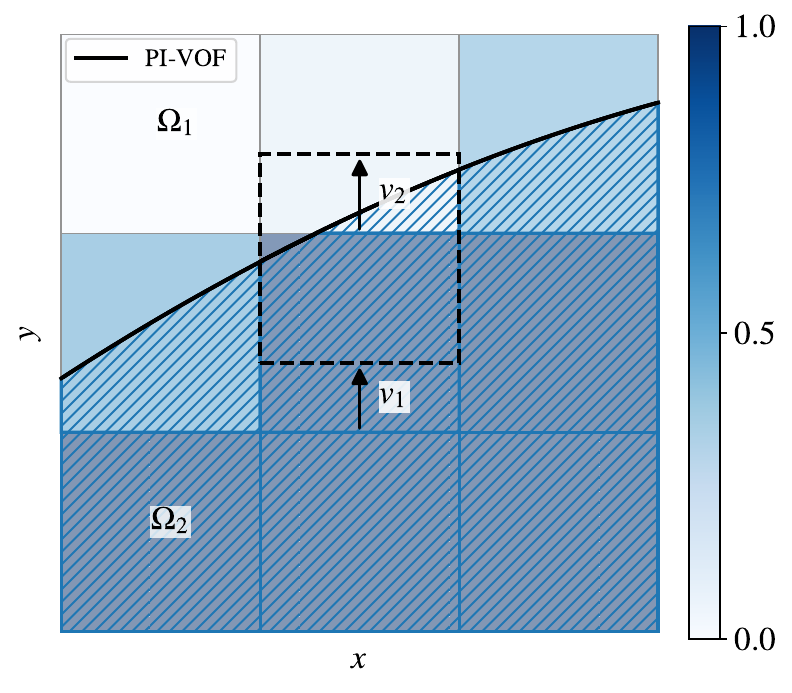}
    \par\smallskip\centering\small (b)
  \end{minipage}

  \caption{Split EI-LE advection mapping using interfaces reconstructed from a sample volume-fraction field using PI-VOF with parametric interfaces. The colour scale denotes the volume fraction, while the hatched and unhatched regions distinguish the phases separated by the PI-VOF reconstructed phase boundaries. 
  (a) Eulerian-implicit $x$-sweep with $u_1$ and $u_2$;
  (b) Lagrangian-explicit $y$-sweep with $v_1$ and $v_2$.}

  \label{fig:eile_mapping}
\end{figure}

Next, a Lagrangian explicit (LE) step in the $y$-direction maps $\Omega_{i,j}^n$ to the target region $\Omega_{\mathrm{target}}$ as
\begin{equation}
\Pi_{y} : \Omega_{i,j}^n \to \Omega_{\mathrm{target}}, \qquad
\begin{cases}
x'' = x, \\
y'' = \gamma_y y + v_1,
\end{cases}
\label{eq:LE_step}
\end{equation}
with $\gamma_y = 1-v_1+v_2$. 
The two directional mappings and the corresponding face-normal
displacements are illustrated schematically in Figure~\ref{fig:eile_mapping}. A Courant number
(CFL$=\max(|u_1|/\Delta x,|u_2|/\Delta x,|v_1|/\Delta y,|v_2|/\Delta y)$) is used to prevent the transported interface from
skipping cells and the time-step $\Delta t_{CFL}$ is restricted to $CFL\leq1$ in the present advection scheme.

\subsection{Heat diffusion and sharp-interface model of solidification}
\label{subsec:heat_transfer}

The solid and liquid regions, $\Omega^s(t)$ and $\Omega^l(t)$, are separated by the reconstructed sharp interface $\Gamma(t)=\partial\Omega^s\cap\partial\Omega^l$. The temperature in each phase satisfies the diffusion equation~\cite{crank1987free}

\begin{equation}
\frac{\partial T}{\partial t}
=
\alpha_p \nabla^2 T +S_{T}
\qquad
\text{in } \Omega^p(t), 
\quad p \in \{s,l\},
\label{eq:heat_eq}
\end{equation}
where $\alpha_p$ denotes the thermal diffusivity of phase $p$ and $S_{T}$ denotes source terms, if any.

\paragraph{Interface conditions}
Across the sharp interface $\Gamma(t)$, the temperature and heat flux satisfy
\begin{align}
T^s &= T^l = T_\Gamma, 
\label{eq:temp_continuity} \\
\left[ k \nabla T \cdot \boldsymbol{n} \right]_\Gamma 
&= \rho L\, V_n ,
\label{eq:stefan_jump}
\end{align}
where $k$ is thermal conductivity, $\rho L$ is latent heat per unit volume,
$\boldsymbol{n}$ is the unit normal pointing into the liquid phase,
and $V_n$ is the normal velocity of the interface.

Equation~\eqref{eq:stefan_jump} represents the classical Stefan condition
expressing local energy balance at the moving boundary.

With the interface normal directed from solid to liquid, the jump operator is defined as $[q]_\Gamma=q^s-q^l$. The Stefan condition determines the normal interface velocity from the one-sided thermal gradients:
\begin{equation}
V_n=\frac{1}{\rho L}\left(k_s\left.\frac{\partial T^s}{\partial n}\right|_\Gamma-k_l\left.\frac{\partial T^l}{\partial n}\right|_\Gamma\right).
\label{eq:stefan_velocity}
\end{equation}
The interfacial temperature is prescribed by the Gibbs--Thomson condition with attachment kinetics,
\begin{equation}
T_\Gamma=T_m-\sigma(\vartheta)\kappa-\beta V_n,
\label{eq:interface_temperature}
\end{equation}
where $\sigma(\vartheta)$ denotes the orientation-dependent capillary coefficient and $\beta$ is the kinetic coefficient. For fourfold anisotropy, the capillary coefficient is expressed as $\sigma(\vartheta)=d_0[1-15\epsilon\cos(4\vartheta)]$.

\paragraph{Temporal discretisation}

The diffusion equation is discretised on the fixed Cartesian grid using a ghost-node finite-difference formulation~\cite{Gibou2003_JSC}. Standard finite-difference stencils are used away from the interface, while stencils intersected by $\Gamma$ are modified to impose the prescribed Dirichlet value $T_\Gamma$. For a stencil connecting nodes $i$ and $i+1$, with the interface located at a fractional distance $\theta\in(0,1)$ from node $i$, the ghost value is obtained by linear extrapolation:
\begin{equation}
T^G_{i+1}=\frac{T^*+(\theta-1)T_i}{\theta},
\label{Eq:GFM_Extrapolation}
\end{equation}
where $T^*$ is the interfacial temperature at the stencil intersection.

The intersection is calculated directly from the reconstructed quadratic segment $\mathbf X(\zeta)$. For horizontal and vertical stencils, respectively, the intersection conditions are $X_y(\zeta)=y_i$ and $X_x(\zeta)=x_i$. Admissible roots within $\zeta\in[0,1]$ provide the interface position and fractional stencil distance. The interfacial temperature is interpolated from the endpoint values as $T^*=T_{\Gamma,0}+\zeta(T_{\Gamma,1}-T_{\Gamma,0})$. The resulting ghost-node contributions are incorporated into the discrete diffusion operator.

The error in the ghost-node solution is separated into the finite-difference discretisation error for a fixed interface and the error introduced by the approximation of the interface geometry. Denoting the reconstructed interface by $\Gamma_h$ and its normal displacement from the exact interface $\Gamma$ by $\delta_n$, the temperature error is expressed as
\begin{equation}
\|T_h-T\|
\lesssim C_1 h^2+C_2\|\delta_n\|,
\label{eq:gfm_error_decomposition}
\end{equation}
where the first contribution represents the underlying spatial discretisation and the second represents sensitivity to interface positioning. This decomposition isolates the effect of quadratic reconstruction on the solution of the sharp-interface diffusion problem.

\subsection{Interface velocity and narrow-band extension}
\label{subsec:velocity_extension}

\begin{figure}[H]
\centering
\begin{minipage}[t]{0.49\linewidth}
\centering
\includegraphics[width=\linewidth]{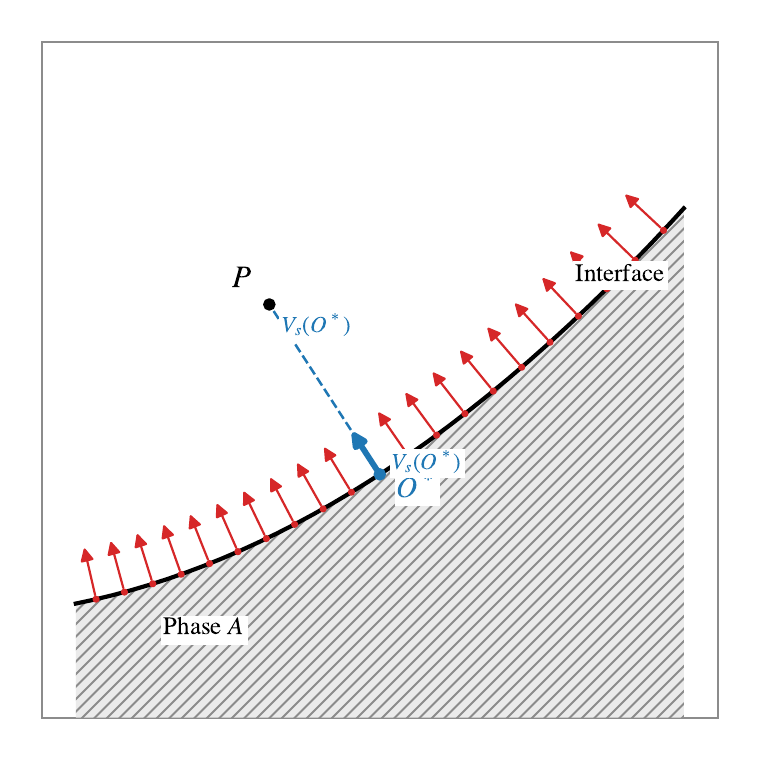}
\par\smallskip\centering\small (a)
\end{minipage}\hfill
\begin{minipage}[t]{0.49\linewidth}
\centering
\includegraphics[width=\linewidth]{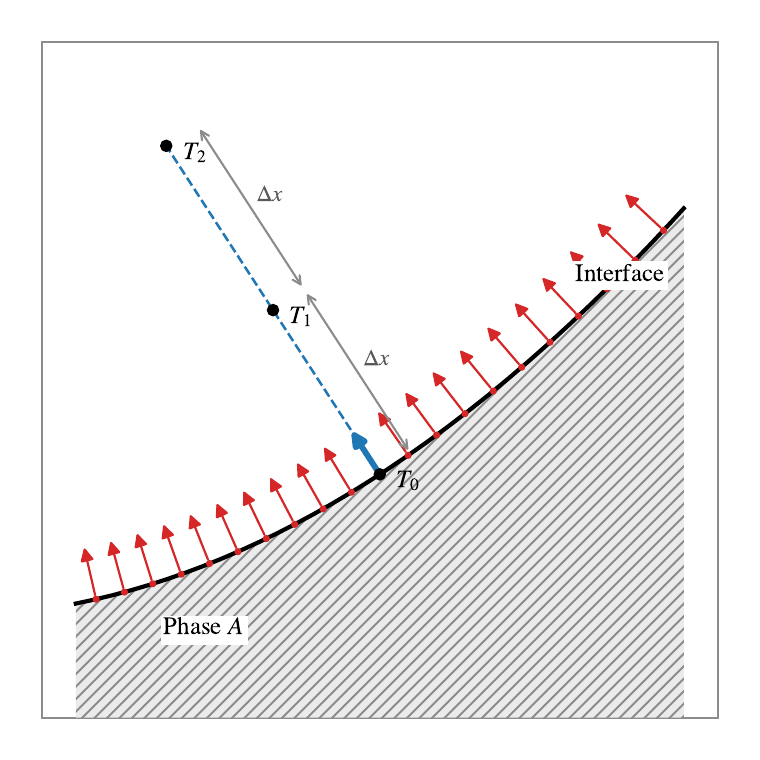}
\par\smallskip\centering\small (b)
\end{minipage}
\caption{Geometric construction for interface propagation: (a) normal projection of a Cartesian point $P$ onto the reconstructed quadratic interface; (b) evaluation of the solid- and liquid-side temperature gradients along the interface normal for the Stefan condition.}
\label{fig:normal_projection}
\end{figure}

The normal interface velocity is obtained from the Stefan condition~\eqref{eq:stefan_jump}. The solid- and liquid-side temperature gradients are evaluated by sampling the temperature field along the reconstructed interface normals, as illustrated in Figure~\ref{fig:normal_projection}(b). The resulting normal velocities are assigned to the reconstructed interface segments and extended to neighbouring Cartesian velocity locations for geometric advection.

The normal velocity prescribed at the interface can be extended into a narrow band by considering the steady normal-extension equation
\begin{equation}
\mathbf{n}\cdot\nabla V_{\mathrm{ext}}=0,
\qquad V_{\mathrm{ext}}|_{\Gamma}=V_n,
\label{eq:extension_pde}
\end{equation}
which preserves the interfacial normal velocity along normal characteristics. Rather than solving this equation over the computational domain, the proposed formulation constructs its local geometric approximation from neighbouring reconstructed parametric curves. For each Cartesian face location $\mathbf{P}$, candidate normal projections are obtained by solving
\begin{equation}
\left[\mathbf{X}(\zeta)-\mathbf{P}\right]\times
\left[\mathbf{n}_0+\zeta(\mathbf{n}_1-\mathbf{n}_0)\right]=0,
\qquad \zeta\in[0,1].
\label{eq:velocity_normal_projection}
\end{equation}
Here, $\mathbf{X}(\zeta)$ denotes the parametric quadratic interface and $\mathbf{n}_0,\mathbf{n}_1$ are its endpoint normals. The resulting cubic equation is solved independently for each candidate segment, with admissible real roots providing local normal projections. The minimum projection distance is evaluated for each segment, followed by distance-based selection over a finite neighbourhood of reconstructed interfaces. When no admissible projection is available, an endpoint-based normal-alignment criterion provides the fallback. The interfacial normal velocity is evaluated at the selected projection and extended along the corresponding normal direction to obtain the Cartesian face-velocity components on the staggered grid used for geometric VOF advection.

\subsection{Curvature evaluation from reconstructed interface}
\label{subsec:curvature}

Curvature is evaluated from the reconstructed interface geometry using a local least-squares fit of interface slopes. Unlike conventional height-function curvature estimation~\cite{Popinet2009}, which evaluates derivatives of heights assembled from volume fractions, the present approach uses the endpoint positions and normals supplied by the quadratic reconstruction.

For each mixed cell, a reference normal is obtained from the reconstructed endpoint normals, and the dominant direction $d$ is selected from its largest absolute component. Interface intersections in the target cell and neighbouring mixed cells provide the fitting coordinates $\xi_i$ along the transverse direction. The corresponding slopes are calculated from the reconstructed normals as
\begin{equation}
h'(\xi_i)=-\frac{n_{\perp,i}}{n_{d,i}},
\label{eq:curvature_slope}
\end{equation}
where $n_d$ and $n_\perp$ are the normal components along the dominant and transverse directions, respectively.

The sampled slopes are approximated by a linear function,
\begin{equation}
h'(\xi)\approx a\xi+b,
\label{eq:slope_fit}
\end{equation}
with coefficients determined through least-squares minimisation:
\begin{equation}
\min_{a,b}\sum_i\left[h'(\xi_i)-(a\xi_i+b)\right]^2.
\label{eq:curvature_lsq}
\end{equation}
This fit corresponds to a local quadratic representation of the interface height. The curvature magnitude at an interface location $\xi$ is then evaluated as

\begin{equation}
\kappa(\xi)=s_\kappa\frac{|a|}{h\left[1+(a\xi+b)^2\right]^{3/2}},
\qquad s_\kappa\in\{-1,1\},
\label{eq:curvature_formula}
\end{equation}
where $h$ is the physical grid spacing and $s_\kappa$ is determined from the local convexity or concavity of the reconstructed interface relative to the phase-oriented normal.

\section{Results and Discussion}

\subsection{Static reconstruction}
\label{sec:static}

The accuracy of static interface reconstruction is assessed using the symmetric-difference area between the reconstructed phase region $\Omega_h$ and the analytical reference region $\Omega_{\mathrm{ref}}$, defined as
\begin{equation}
E_{\Delta} = |\Omega_h \triangle \Omega_{\mathrm{ref}}|,
\end{equation}
where $\triangle$ denotes the symmetric difference and $|\cdot|$ denotes the enclosed area. This measure accounts for both over- and under-enclosed regions of a particular phase. The reconstruction evaluates discrete curvature value in each mixed interfacial cell. Hence, curvature errors are evaluated against the analytical curvature of the corresponding interface. The interface length weighed and normalized $L_1$-type curvature error is defined as
\begin{equation}
E_{\kappa} = \frac{1}{L_{\mathrm{ref}}\kappa_{\mathrm{ref,max}}}\sum_{i=1}^{M}L_i
\left|
\,|\kappa_{h,i}|-|\kappa_{\mathrm{ref}}(\mathbf{x}_i)|
\right|
\label{eq:static_curvature_error}
\end{equation}
Here, $M$ is the number of mixed interface cells, $L_i$ is the reconstructed interface length in cell $i$, and $\kappa_{h,i}$ is the corresponding discrete curvature. The quantities $L_{\mathrm{ref}}$ and $\kappa_{\mathrm{ref,max}}$ denote the total analytical interface length and maximum absolute analytical curvature, respectively.

\begin{figure}[H]
\centering
\begin{minipage}[t]{0.48\linewidth}
\includegraphics[trim={0cm 0cm 0cm 0cm},clip,scale=0.5]{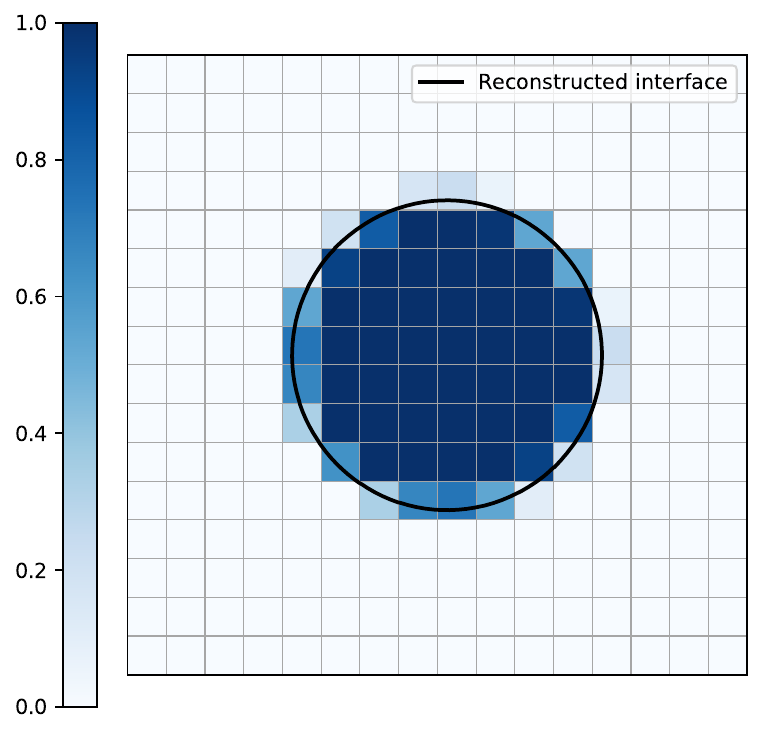}
\par\smallskip\centering\small (a)
\end{minipage}\hfill
\begin{minipage}[t]{0.48\linewidth}
\includegraphics[scale=0.45]{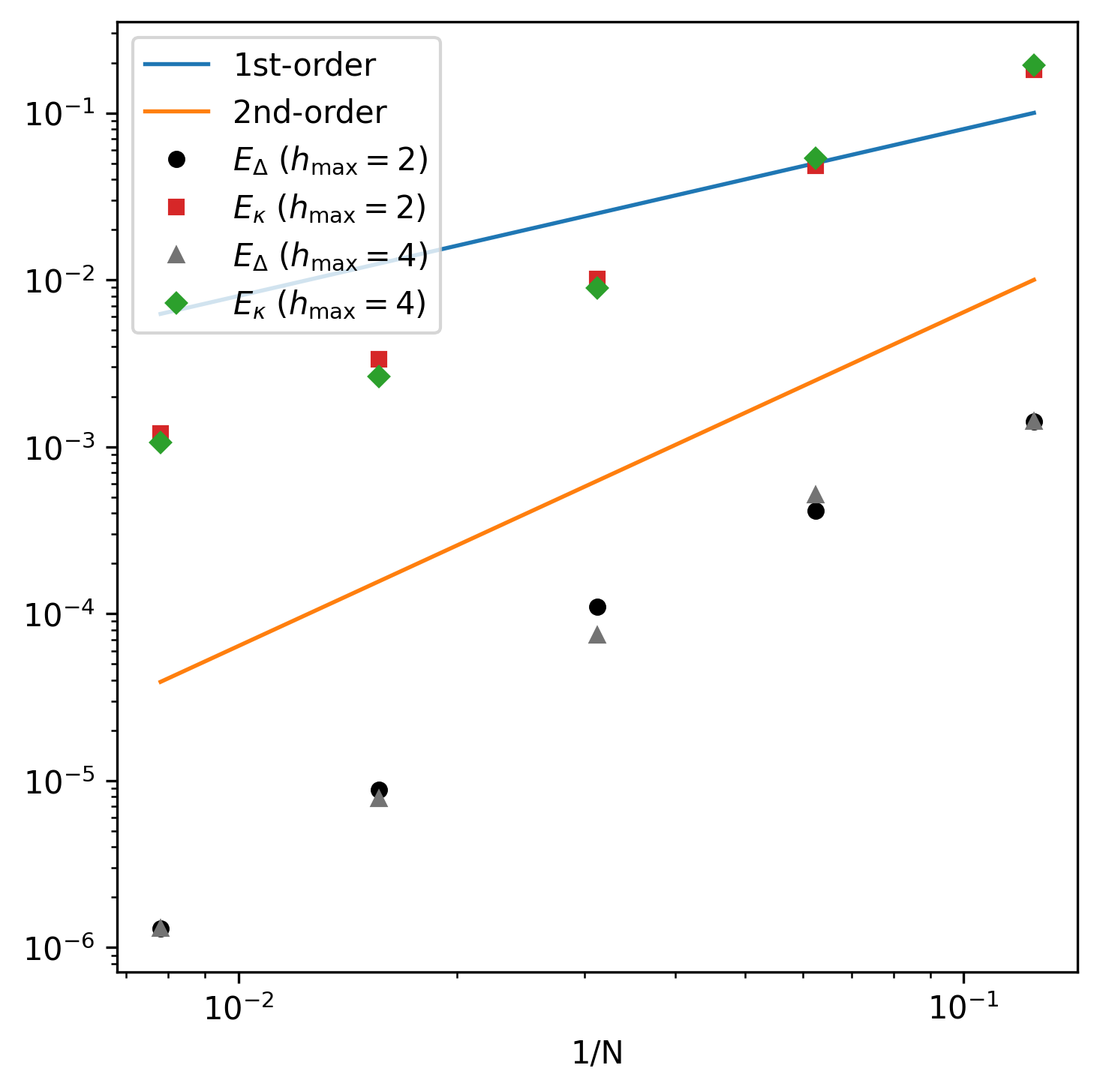}
\par\smallskip\centering\small (b)
\end{minipage}

\caption{Static reconstruction of circular interfaces with randomly perturbed centre positions: (a) representative volume-fraction field with the quadratic reconstructed interface (black curve) at $N=16$; (b) symmetric-difference area error $E_{\Delta}$ and  curvature error $E_{\kappa}$ under mesh refinement for $h_{\max}=2$ and $h_{\max}=4$. First- and second-order reference slopes are shown for comparison.}
\label{fig:circle}
\end{figure}

A circular interface of radius $R=0.25$ is considered in the unit domain $\Omega=(0,1)^2$. The circle centre is randomly positioned within a disk of radius $0.05$ centred at $(0.5,0.5)$ to reduce grid-alignment bias. The reconstruction errors are evaluated over 100 samples at each resolution. The analytical curvature is $\kappa=1/R$. Figure~\ref{fig:circle} presents a representative volume-fraction field and the convergence of both error measures. Results are presented for maximum height-function search extents of $h_{\max}=2$ and $h_{\max}=4$ cells. The symmetric-difference error converges at a rate above second order on the finer meshes with average order approximately $2.5$ for both cases, while the normalized curvature error exhibits average convergence orders of approximately $1.8$ and $1.9$ for $h_{\max}=2$ and $h_{\max}=4$, respectively. The smaller search extent, $h_{\max}=2$, is adopted for all subsequent simulations.

\subsection{Advection: Reversible vortex}
\label{sec:vortex}
Before considering interface evolution, we validate the implementation of split advection with the proposed locally reconstructed PI-VOF.
A circle of radius $0.15$ centered at $(0.5,0.75)$ is advected in $\Omega=(0,1)^2$ using the solenoidal velocity field~\cite{Scardovelli2003}
\begin{equation}
u = \sin^2(\pi x)\sin(2\pi y)\cos(\pi t/T),\qquad 
v = -\sin(2\pi x)\sin^2(\pi y)\cos(\pi t/T).
\end{equation}

\begin{figure}[H]
\centering
\begin{minipage}[t]{0.48\linewidth}
\centering
\includegraphics[width=\linewidth,trim={0cm 0cm 10.5cm 0cm},clip]{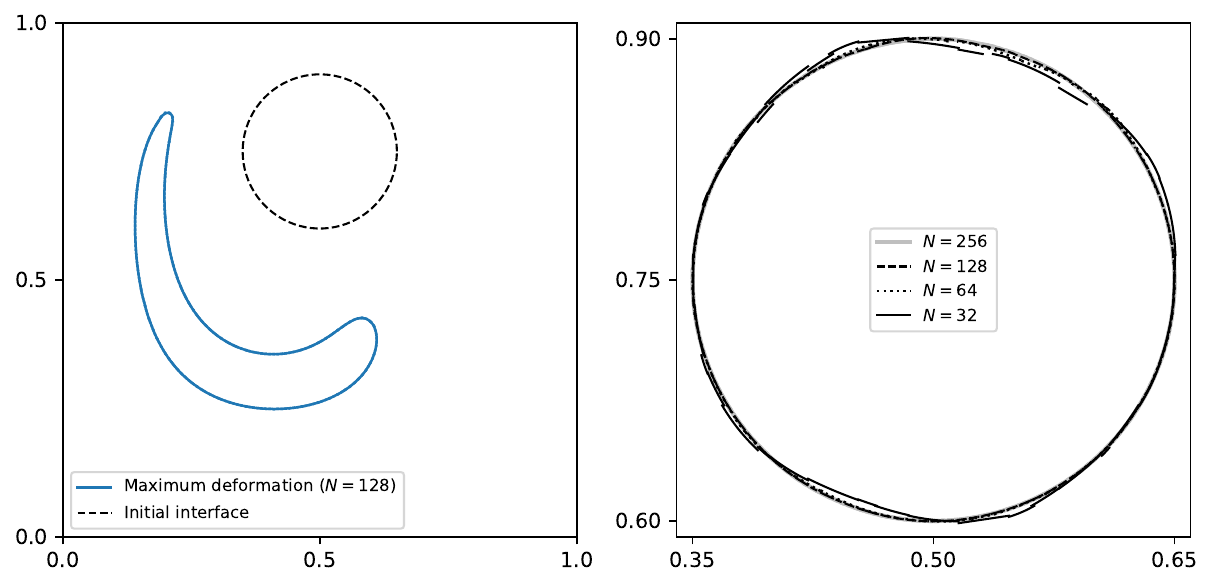}
\par\smallskip\centering\small (a)
\end{minipage}\hfill
\begin{minipage}[t]{0.48\linewidth}
\centering
\includegraphics[width=\linewidth,trim={10.5cm 0cm 0cm 0cm},clip]{sections/vortex_T2.pdf}
\par\smallskip\centering\small (b)
\end{minipage}
\caption{Reversible-vortex advection using PI-VOF at $T=2$: (a) reconstructed interfaces at maximum deformation ($t=T/2$) and after flow reversal ($t=T$); (b) reconstructed interfaces at $t=T$ for different grid resolutions.}
\label{fig:vortex}
\end{figure}
The prescribed velocity field is divergence-free, and the interface reaches maximum deformation at $t=T/2$, followed by reversal to its initial configuration at $t=T$. Figure~\ref{fig:vortex} illustrates the interface deformation and recovery using PI-VOF, including the effect of grid refinement on the final reconstructed interface. Simulations are performed for $T=1$ and $T=2$ with CFL$=1.0$. The accuracy of the reconstruction--advection combination is assessed using the symmetric-difference error, the cellwise volume-fraction return error, and the relative mass error. Following Scardovelli and Zaleski~\cite{Scardovelli2003}, the cellwise volume-fraction error is defined as
\begin{equation}
E_1=h^2\sum_{i,j}|f_{i,j}(T)-f_{i,j}(0)|,
\end{equation}
where $h=1/N$. The symmetric-difference error $E_{\Delta}$ measures the area of the symmetric difference between the reconstructed phase regions at $t=0$ and $t=T$. Mass conservation is quantified using
\begin{equation}
E_m=\frac{|M(T)-M(0)|}{M(0)},\qquad
M(t)=h^2\sum_{i,j}f_{i,j}(t).
\end{equation}

\begingroup
\sisetup{round-mode=figures,round-precision=3,exponent-mode=scientific}
\begin{table}[H]
\centering
\small
\setlength{\tabcolsep}{2pt}
\caption{Reversible-vortex errors for $T=1$ and $T=2$ at CFL$=1.0$ using PI-VOF. The present results report the cellwise volume-fraction error $E_1$, symmetric-difference area $E_{\Delta}$, and relative mass error $E_m$.}

\label{tab:vortex_errors}
\begin{tabular}{@{} >{\raggedright\arraybackslash}p{0.18\linewidth} S S S S S @{}}
\hline
{N / error} & {$16$} & {$32$} & {$64$} & {$128$} & {$256$} \\
\hline

\multicolumn{6}{l}{\textit{(a) Reversal period $T=1$}} \\
\hline
{PI-VOF, $E_1$} & \num{0.00241263503648915} & \num{0.0006419641375759495} & \num{0.0001508674172716058} & \num{3.6280335867071085e-05} & \num{9.232697542093376e-06} \\
{PI-VOF, $E_{\Delta}$} & \num{0.0026039497516176426} & \num{0.0006813053673300457} & \num{0.00015604722858284518} & \num{3.756175694886958e-05} & \num{9.489606770000001e-06} \\
{PI-VOF, $E_m$} & \num{1.9633053589280963e-16} & \num{1.963305358928097e-16} & \num{1.0277314562380908e-11} & \num{3.6859094808516094e-12} & \num{9.61234303731196e-13} \\
\hline
\multicolumn{6}{l}{\textit{(b) Reversal period $T=2$}} \\
\hline
{PI-VOF, $E_1$ }& \num{0.008848284728912262} & \num{0.0021072461092302027} & \num{0.0004421357968234196} & \num{0.00012099418451556139} & \num{2.508161827977246e-05} \\
{PI-VOF, $E_{\Delta}$} & \num{0.008962272002646364} & \num{0.0021414234713637} & \num{0.0004488125700941711} & \num{0.00012171683710561116} & \num{2.5264314260520663e-05} \\
{PI-VOF, $E_m$} & \num{0.0001023925764552941} & \num{0.0} & \num{2.03830362363915e-12} & \num{1.157074013284274e-11} & \num{8.182075083332842e-12} \\

{ELVIRA~\cite{Scardovelli2003}, $E_1$} &{-} & \num{2.52e-3} & \num{6.46e-4} & \num{1.45e-4} & {-} \\
{Quadratic fit~\cite{Scardovelli2003}, $E_1$} & {-} & \num{1.88e-3} & \num{4.42e-4} & \num{9.36e-5} & {-} \\
{Quadratic fit - continuity~\cite{Scardovelli2003}, $E_1$} & {-} & \num{1.09e-3} & \num{2.80e-4} & \num{5.72e-5} & {-} \\

\hline
\end{tabular}
\end{table}
\endgroup

Table~\ref{tab:vortex_errors} reports the errors for both reversal periods, together with representative results from the split Eulerian-Lagrangian schemes of Scardovelli and Zaleski~\cite{Scardovelli2003}. The present $E_1$ errors are comparable in magnitude to those reported for these schemes, although continuity-constrained reconstructions yield smaller absolute errors. Both $E_1$ and $E_{\Delta}$ exhibit approximately second-order convergence, with average observed orders of $2.01$ and $2.03$ for $T=1$, and $2.12$ for both measures at $T=2$. The relative mass error remains below $1.16\times10^{-11}$ except on the coarsest $T=2$ grid, where $E_m=1.02\times10^{-4}$.
These results demonstrate that, for the prescribed divergence-free flow, the PI-VOF reconstruction and split-advection combination retains volume conservation to numerical tolerance on the refined grids while achieving approximately second-order accuracy. This establishes the conservative transport treatment suitable for use with the proposed geometric normal-velocity extension in the subsequent moving-interface and Stefan problems.

\subsection{Prescribed normal-velocity evolution: Expanding circle}
\label{sec:expanding_circle}

\begin{figure}[H]
\centering
\begin{minipage}[t]{0.48\linewidth}
\centering
\includegraphics[width=\linewidth]{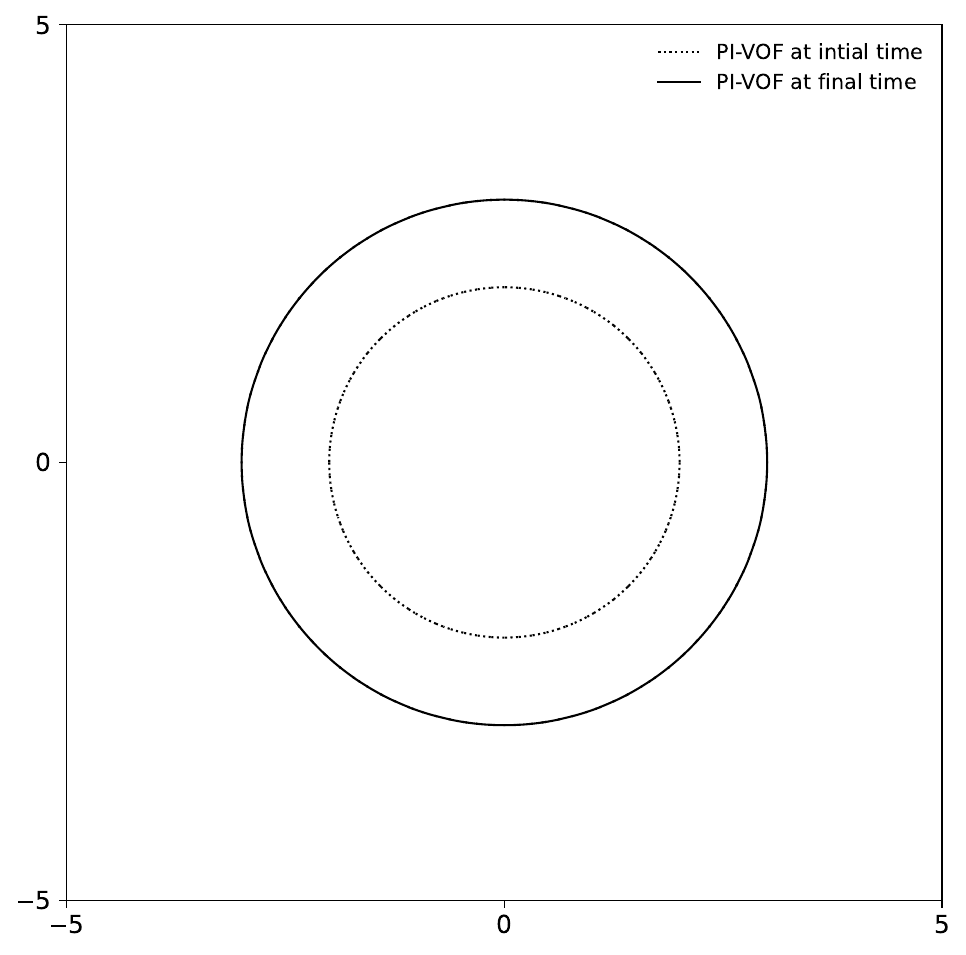}
\par\smallskip{\small (a)}
\end{minipage}\hfill
\begin{minipage}[t]{0.48\linewidth}
\centering
\includegraphics[width=\linewidth]{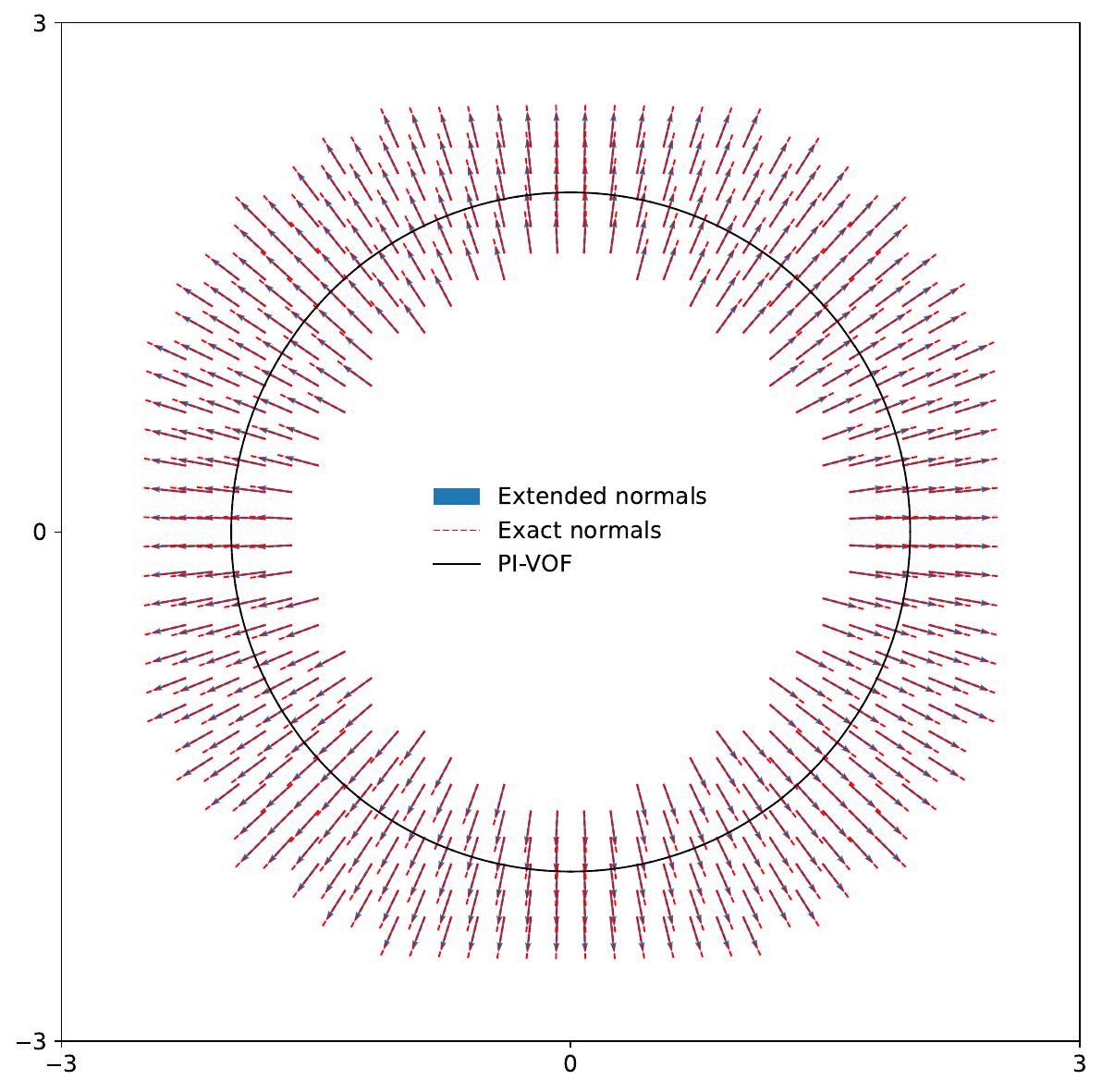}
\par\smallskip{\small (b)}
\end{minipage}
\caption{Prescribed expansion of a circular interface using PI-VOF at $N=64$: (a) initial and final reconstructed interfaces at $t=0$ and $t=1$; (b) comparison of the geometrically extended normal vectors in the narrow band with the analytical radial normals.}
\label{fig:expanding_circle}
\end{figure}

To validate interface evolution under a prescribed normal velocity, a circular interface of initial radius $R_0=2$ is expanded with a constant outward normal speed $V_n=1$ in a $10\times10$ domain with CFL-limited (CFL$=0.5$) time steps. Unlike the preceding divergence-free advection test, the prescribed motion changes the enclosed phase volume. The analytical radius is
\begin{equation}
R(t)=R_0+V_nt=2+t,
\end{equation}
The normal velocity prescribed at the reconstructed interface is extended into a narrow band to evolve the volume-fraction field. Figure~\ref{fig:expanding_circle} shows the initial and final PI-VOF interfaces at $N=64$, together with the extended normal vectors and their analytical counterparts.

In addition to the symmetric-difference error $E_{\Delta}$, the accuracy of the enclosed phase area is evaluated through the equivalent-area radius and its absolute error,
\begin{equation}
R_A(t)=\sqrt{\frac{h^2\sum_{i,j}f_{i,j}(t)}{\pi}},\qquad E_R=|R_A(t)-R(t)|.
\end{equation}

Table~\ref{tab:expanding_circle} reports the errors at $t_f=1$. The equivalent-area radius error exhibits an average convergence order of $2.05$, while the symmetric-difference error converges with an average order of $1.85$. These results confirm that the PI-VOF reconstruction and geometric normal-velocity extension reproduce the prescribed interface growth with approximately second-order accuracy, including the corresponding change in enclosed phase area.

\begingroup
\sisetup{round-mode=figures,round-precision=3,exponent-mode=scientific}
\begin{table}[H]
\centering
\small
\caption{Convergence of equivalent-area radius error $E_R$ and symmetric-difference area error $E_{\Delta}$ for prescribed circular expansion at $t_f=1$.}
\label{tab:expanding_circle}
\begin{tabular}{@{} c S c S c @{}}
\hline
{$N$} & {$E_R$} & {EOC} & {$E_{\Delta}$} & {EOC} \\
\hline
16  & \num{0.00594967480639275}  & --   & \num{0.168198347091675}   & --   \\
32  & \num{0.00128669117384606}  & 2.21 & \num{0.04584901034832}    & 1.88 \\
64  & \num{0.000375139442745365} & 1.78 & \num{0.0117536634206772}  & 1.96 \\
128 & \num{8.29327959621651e-05} & 2.18 & \num{0.00357858370989561} & 1.72 \\
\hline
\end{tabular}
\end{table}
\endgroup

\subsection{Curvature-driven interface evolution}
\label{sec:curvature_evolution}

The preceding test considered interface evolution under a constant prescribed normal velocity. Here, the normal velocity is a function of local interface curvature. Following curvature-dependent propagation models used in level-set methods~\cite{Aslam1996}, the outward normal velocity is prescribed as
\begin{equation}
V_n=1-\sigma\kappa,\qquad \sigma=0.1,
\end{equation}
where $\kappa$ denotes the local interface curvature. For an initially circular interface of radius $R_0=0.4$ located near center of a square domain $\Omega=[-1,1]^2$, the curvature remains spatially uniform, $\kappa=1/R(t)$, and the analytical radius satisfies
\begin{equation}
\frac{dR}{dt}=1-\frac{0.1}{R},\qquad
t=R(t)-R_0+0.1\ln\left(\frac{R(t)-0.1}{R_0-0.1}\right).
\end{equation}

\begin{figure}[H]
\centering
\begin{minipage}[t]{0.48\linewidth}
\centering
\includegraphics[width=\linewidth]{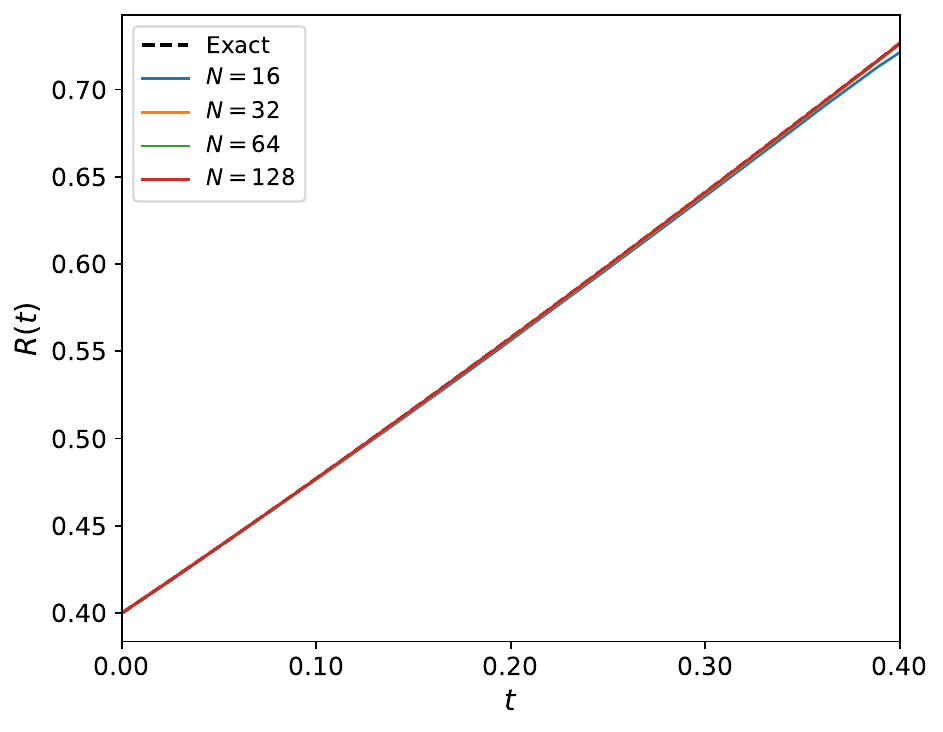}
\par\smallskip\centering\small (a)
\end{minipage}\hfill
\begin{minipage}[t]{0.48\linewidth}
\centering
\includegraphics[width=\linewidth]{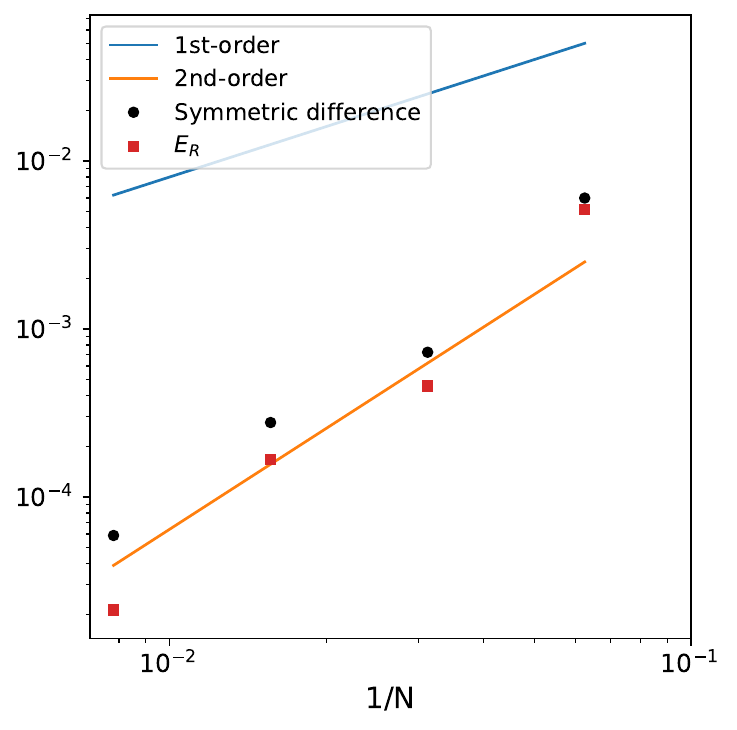}
\par\smallskip\centering\small (b)
\end{minipage}
\caption{Curvature-dependent expansion of a circular interface using PI-VOF with $V_n=1-0.1\kappa$: (a) numerical radius evolution for $N=16,32,64,128$ compared with the exact radius vs time; (b) symmetric-difference error and average radius error at $t_f=0.4$ under mesh refinement.}
\label{fig:curvature_evolution}
\end{figure}

Simulations are performed on grids of $N=16,32,64,128$ up to $t_f=0.4$. 
The time step is selected using a curvature-dependent stability restriction, following the linearised stability~\cite{Aslam1996}, together with an interface CFL constraint:
\begin{equation}
\Delta t=\min\left(\Delta t_{\mathrm{CFL}},\frac{C_{\kappa}h^2}{2h+2\sigma}\right)
\end{equation}
Here, $C_{\kappa}=0.8$ is the adopted stability factor, while the additional interface CFL restriction ($\leq0.5$) controls the geometric advection step.
 Figure~\ref{fig:curvature_evolution}(a) compares the numerical radius evolution with the analytical solution, while Fig.~\ref{fig:curvature_evolution}(b) reports the symmetric-difference error at $t_f=0.4$ under mesh refinement. The results demonstrate radius convergence towards the analytical curvature-dependent growth and verify that the symmetric-difference convergence approaches near-second order at finer resolutions.

\subsection{Stefan problem: Interface growth without curvature effects}
\label{sec:frank}

The Frank circle benchmark validates the coupling among the thermal solver, the Stefan interface condition, and the geometric normal-velocity extension in PI-VOF. Unlike the preceding tests with prescribed normal velocities, the interface velocity is now determined by the temperature gradients on either side of the interface. 
The two-dimensional Frank circle problem is solved in $\Omega=[-5,5]^2$ using the analytical similarity solution~\cite{Gibou2003_JSC}. Let $\rho=\sqrt{x^2+y^2}$ denote the radial distance from the centre. The analytical interface radius is

\begin{equation}
R(t)=s_0\sqrt{t}
\end{equation}
with temperature
\begin{equation}
T(\rho,t)=
\begin{cases}
0, & \rho<R(t),\\[2mm]
T_\infty\left[
1-\dfrac{E_1\!\left(\rho^2/(4t)\right)}
{E_1\!\left(s_0^2/4\right)}
\right], & \rho\ge R(t),
\end{cases}
\end{equation}

where $\operatorname{E}_1$ is the exponential integral, and unit thermal diffusivity is assumed. The parameters are $T_\infty=-0.5$ and $s_0=1.56$. The simulation starts at $t=1$, corresponding to $R(1)=1.56$, and continues until $t=1.5$. The initial temperature field is prescribed from the analytical solution. The time step is selected proportional to the smallest cuts allowed, $0.5(\theta h)$, where $\theta$ is the minimum-cut parameter used in the numerical discretisation. The temperature equation is advanced using the Crank-Nicolson
scheme. The analytical temperature is imposed on the outer boundary,
while the zero interface temperature is imposed sharply through the
ghost-fluid discretization.

\begin{figure}[H]
\centering
\begin{minipage}[t]{0.48\linewidth}
\centering
\includegraphics[width=\linewidth]{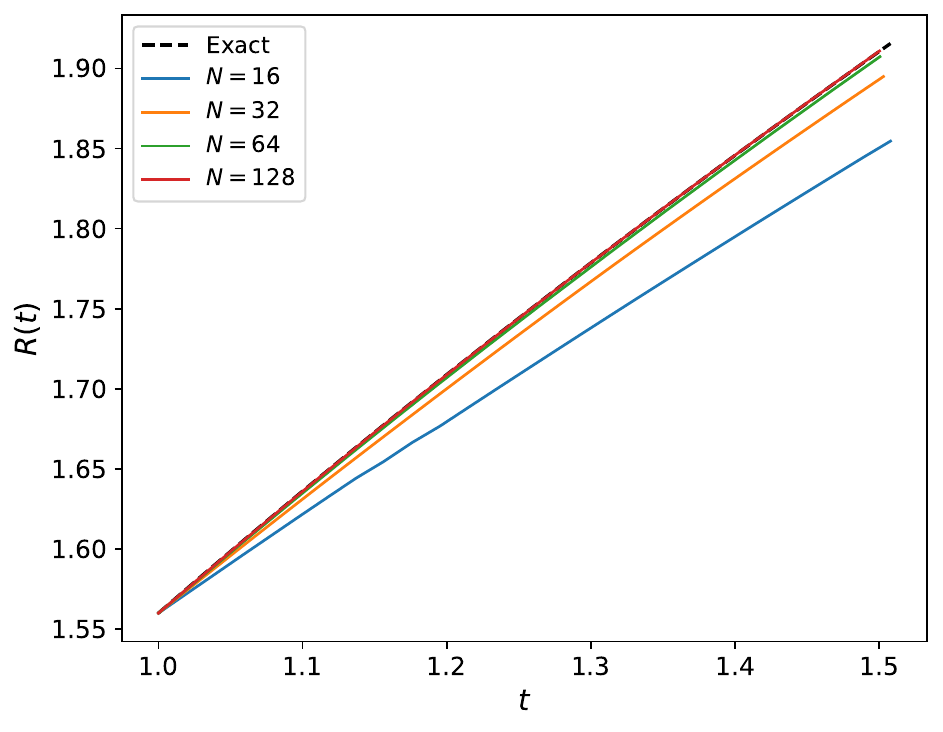}
\par\smallskip\centering\small (a)
\end{minipage}\hfill
\begin{minipage}[t]{0.48\linewidth}
\centering
\includegraphics[width=\linewidth]{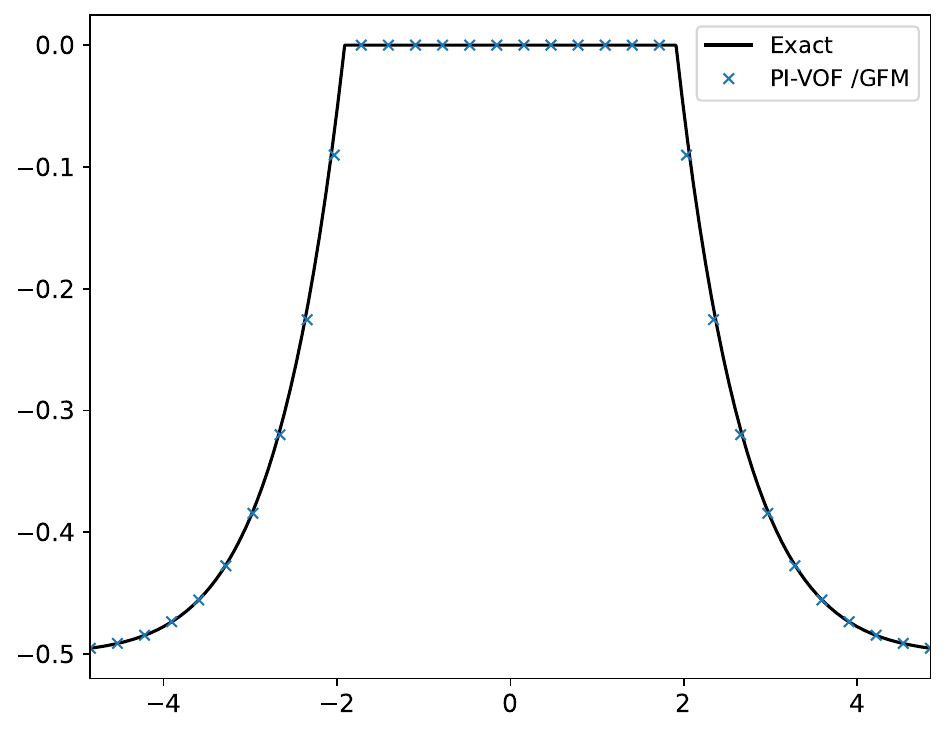}
\par\smallskip\centering\small (b)
\end{minipage}
\caption{Frank circle benchmark using PI-VOF: (a) numerical radius evolution compared with the exact solution; (b) centreline temperature
at $t=1.5$ for $N=32$ compared with the analytical solution.}

\label{fig:frank}
\end{figure}

Figure~\ref{fig:frank}(a) compares the numerical equivalent-area radius with the analytical evolution, while Fig.~\ref{fig:frank}(b) compares the numerical and analytical temperature profiles. Table~\ref{tab:frank} reports the symmetric-difference error, equivalent-area radius error, and temperature $L_2$ error at the final time. The geometric and radius errors decrease monotonically under refinement, while the temperature field exhibits approximately second-order convergence. These results validate the coupling of the ghost-fluid thermal discretization, Stefan interface condition, and geometric normal-velocity extension with the local parametric interface representation in PI-VOF.

\begingroup
\sisetup{round-mode=figures,round-precision=3,exponent-mode=scientific}
\begin{table}[H]
\centering
\small
\caption{Convergence of symmetric-difference area, mean-radius error, and temperature $L_2$ error for the Frank circle benchmark at $t=1.5$.}
\label{tab:frank}
\begin{tabular}{@{} c S c S c S c @{}}
\hline
{$N$} & {$E_\Delta$} & {EOC} & {$E_R$} & {EOC} & {$L_2$} & {EOC} \\
\hline
16  & \num{2.637744e+00} & --    & \num{6.017138e-02} & --    & \num{3.594697e-02} & -- \\
32  & \num{2.085020e-01} & 3.661 & \num{1.744865e-02} & 1.786 & \num{2.856772e-03} & 3.653 \\
64  & \num{4.367135e-02} & 2.255 & \num{3.567016e-03} & 2.290 & \num{7.073590e-04} & 2.014 \\
128 & \num{1.983511e-02} & 1.139 & \num{1.542381e-04} & 4.531 & \num{1.249790e-04} & 2.501 \\
\hline
\end{tabular}
\end{table}
\endgroup

\subsection{Stefan problem: Interface growth with curvature and attachment-kinetic effects}
\label{sec:curvature}

The preceding Frank circle test considered interface evolution in the absence of curvature or kinetic effects. The present benchmark introduces these contributions through the Gibbs-Thomson condition with kinetic undercooling,
\begin{equation}
T_\Gamma=-\sigma\kappa-\beta V_n,
\end{equation}
where $\sigma$ and $\beta$ denote the capillary and kinetic coefficients, respectively. A manufactured radially symmetric solution~\cite{schmidt1994} is considered on
$\Omega=[-2,2]^2$, with initial radius $R_0=0.5$ and
\begin{equation}
R(t)=\sqrt{R_0^2+2\lambda t},\qquad
V_n(t)=\frac{\lambda}{R(t)},\qquad \lambda=0.5.
\end{equation}
For a circular interface, $\kappa=1/R(t)$, and hence
\begin{equation}
T_\Gamma(t)
=-\frac{\eta(1+\lambda)}{R(t)}.
\end{equation}
With $\rho=\sqrt{x^2+y^2}$, the corresponding temperature is
\begin{equation}
T(\rho,t)=
\begin{cases}
T_\Gamma(t), & \rho\le R(t),\\[2mm]
T_\Gamma(t)
-\lambda e^{\lambda/2}
\displaystyle\int_1^{\rho/R(t)}
\frac{e^{-\lambda y^2/2}}{y}\,dy,
& \rho>R(t).
\end{cases}
\end{equation}
Here, $\eta=10^{-3}$, with $\sigma=\beta=\eta$. The manufactured heat source is $S_T(t)=\eta\lambda(\lambda+1)/R^3(t)$, with unit thermal diffusivity and latent heat. The initial temperature and outer Dirichlet boundary conditions are prescribed from Eq.~(43), and the solution is advanced to $t_f=0.5$. The temperature field continues to satisfy the heat equations and Stefan balance described in Section~\ref{sec:frank}. The interface curvature is evaluated from the PI-VOF reconstruction, while the normal velocity is obtained from the Stefan condition and geometrically extended into the narrow band. The time step is selected proportional to $\theta h$, as in the Frank circle benchmark. At each mixed cell, the Stefan velocity and the kinetic interface
temperature are coupled through a weighted fixed-point correction.
The correction uses the relaxation factor
$\omega=(1+3\beta C)^{-1}$, where
$C=\alpha/(\mathcal{L}h)$ and $\mathcal{L}$ is the latent heat.

\begin{figure}[H]
\centering
\begin{minipage}[t]{0.48\linewidth}
\centering
\includegraphics[width=\linewidth]{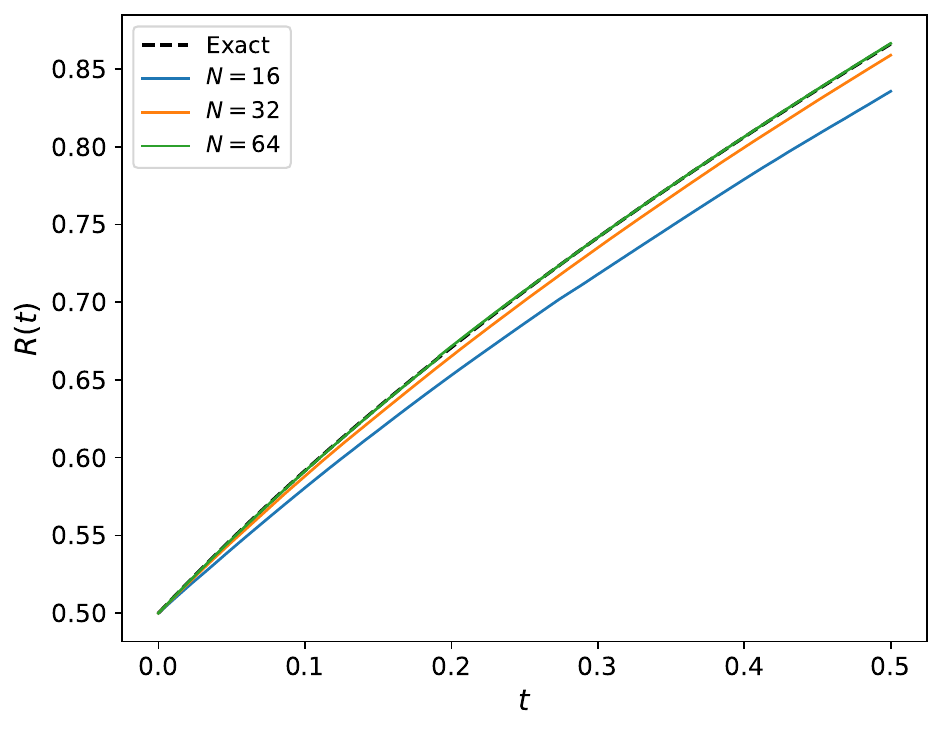}
\par\smallskip\centering\small (a)
\end{minipage}\hfill
\begin{minipage}[t]{0.48\linewidth}
\centering
\includegraphics[width=\linewidth]{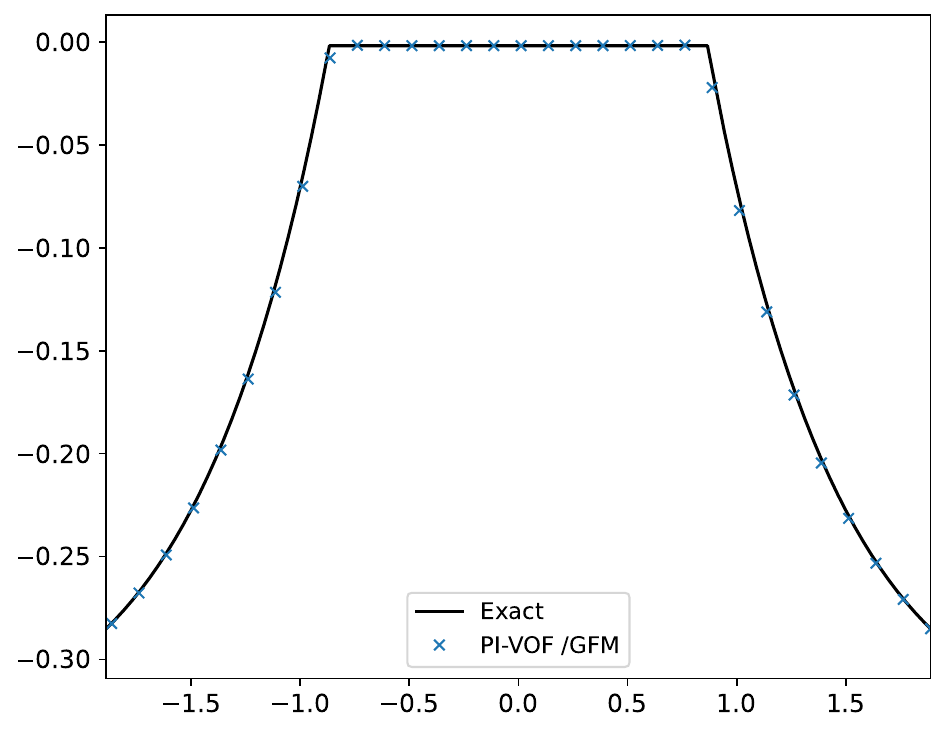}
\par\smallskip\centering\small (b)
\end{minipage}
\caption{Stefan interface evolution with curvature and kinetic effects using PI-VOF: (a) numerical mean-radius evolution compared with the analytical solution; (b) numerical and analytical temperature profiles at $N=32$.}
\label{fig:curvature}
\end{figure}

Figure~\ref{fig:curvature}(a) compares the numerical equivalent-area radius with the analytical evolution, while Fig.~\ref{fig:curvature}(b) compares the numerical and analytical temperature profiles. Table~\ref{tab:curvature} reports the corresponding convergence results. The temperature $L_2$ error exhibits observed orders of $1.81$ and $2.38$, indicating approximately second-order thermal accuracy. Both the equivalent-area radius and symmetric-difference errors decrease with refinement, demonstrating convergent interface evolution that incorporates curvature and kinetic effects. These results support the combined implementation of local parametric reconstruction, curvature evaluation, ghost-fluid interface conditions, Stefan velocity calculation, and geometric normal-velocity extension in PI-VOF.

\begingroup
\sisetup{round-mode=figures,round-precision=3,exponent-mode=scientific}
\begin{table}[H]
\centering
\small
\caption{Convergence of symmetric-difference area, mean-radius error, and temperature $L_2$ error for the Stefan benchmark with curvature and kinetic effects.}
\label{tab:curvature}
\begin{tabular}{@{} c S c S c S c @{}}
\hline
{$N$} & {$E_\Delta$} & {EOC} & {$E_R$} & {EOC} & {$L_2$} & {EOC} \\
\hline
16 & \num{1.631866e-01} & --    & \num{3.038720e-02} & --    & \num{6.456562e-03} & -- \\
32 & \num{4.247915e-02} & 1.942 & \num{7.171019e-03} & 2.083 & \num{1.842201e-03} & 1.809 \\
64 & \num{1.448669e-02} & 1.552 & \num{3.618798e-04} & 4.309 & \num{3.548458e-04} & 2.376 \\
\hline
\end{tabular}
\end{table}
\endgroup

\subsection{Stefan problem: Interface growth with anisotropic curvature effects}
\label{sec:anisotropic}
We next consider dendritic solidification under a dimensionless undercooling of $\Delta=-0.55$, incorporating anisotropic capillary effects.
The interfacial temperature is modified through an anisotropic Gibbs--Thomson relation,
\begin{equation}
T_{\Gamma} = -\kappa \Gamma 
\left(1 - \delta \cos\big(a(\phi - \theta)\big)\right),
\label{eq:anisotropicGT}
\end{equation}
where $\kappa$ is curvature, $\Gamma$ is the capillary length, $\delta=15\epsilon$ is the fourfold capillary anisotropy coefficient ($a=4$), $\phi$ is the local interface-normal orientation, and $\theta$ is the prescribed crystallographic orientation. Anisotropic capillarity leads to morphological selection of dendritic growth directions in preferential crystallographic directions. The simulations use unit thermal diffusivity and latent heat. 
The dendrite-tip velocity is compared with the microscopic-solvability-theory prediction reported by Kim et al.~\cite{Kim20000} for the same undercooling.

\begin{figure}[H]
\centering
\begin{minipage}[b]{0.48\textwidth}
\includegraphics[width=\textwidth,trim={0 0 0 0.7cm},clip]{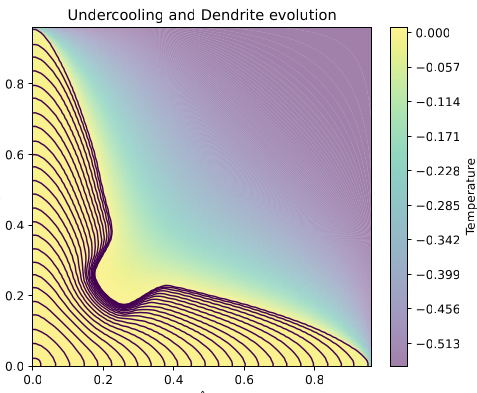}
\par\smallskip\centering\small (a)
\end{minipage}
\begin{minipage}[b]{0.48\textwidth}
\includegraphics[width=\textwidth,trim={0 0 0 1.2cm},clip]{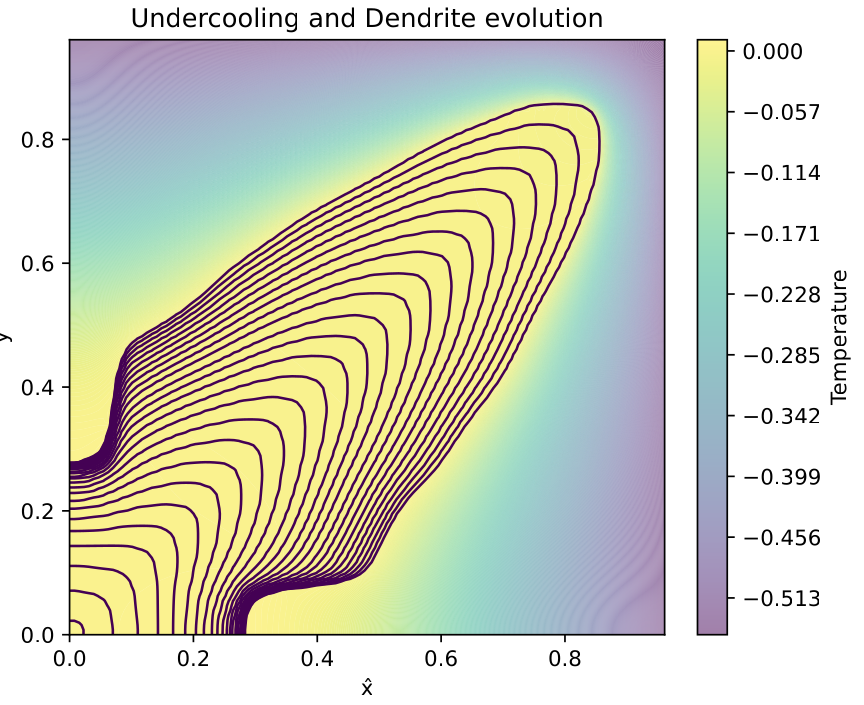}
\par\smallskip\centering\small (b)
\end{minipage}
\caption{Dendritic evolution under anisotropic solidification for two orientations: (a) $\theta=0^\circ$; (b) $\theta=45^\circ$. The spatial coordinates are normalised by the domain length $L$, with $x/L,y/L\in[0,1]$.}
\label{fig:anisotropic_orient_temp}
\end{figure}
\begin{figure}[H]
\centering
\begin{minipage}[b]{0.48\textwidth}
\includegraphics[width=\textwidth]{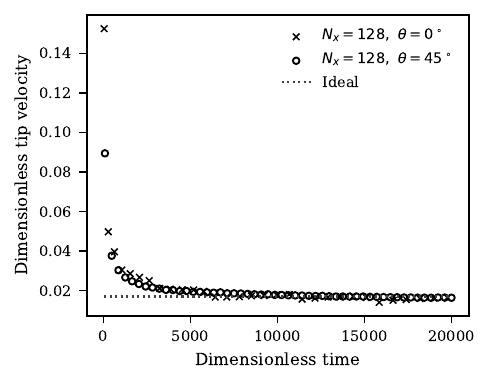}
\end{minipage}

\caption{Evolution of the dimensionless dendrite-tip velocity for $\theta=0^\circ$ and $45^\circ$, compared with the microscopic-solvability-theory reference reported by Kim et al.~\cite{Kim20000}.}

\label{fig:anisotropic_orient}
\end{figure}

The numerical parameters are $\Gamma=d_0=0.5$, $\epsilon=0.05$, and $\beta=0$. The computation uses a quadrant of side length $L=800$, with symmetry conditions on the coordinate axes and $T=\Delta$ on the outer boundaries. For $N=128$, the initial circular seed has radius $R_0=30$, with $T(\mathbf{x},0)=-d_0/R_0$ inside the seed and $T(\mathbf{x},0)=\Delta$ elsewhere. The time step is $\Delta t=C_{\mathrm{I}}h/\max_{\Gamma}|V_n|$, where $h=L/N$ and $C_{\mathrm{I}}=0.1$. The plots in Figure ~\ref{fig:anisotropic_orient_temp} depict the evolution of dendrites in different orientations. This case evaluates orientation-dependent curvature, the stability of anisotropic Gibbs-Thomson enforcement, and the robustness of geometric velocity extension. 
The interface evolution approaches a steady propagation regime, consistent with classical dendritic growth behaviour, as shown in Figure ~\ref{fig:anisotropic_orient}. 
The dendrite-tip position $h_{\mathrm{tip}}(t)$ is determined along the prescribed growth direction $\theta$ using a local height-function construction. Three adjacent rows or columns are used, depending on the growth orientation, with an adaptive search in the transverse direction to locate the tip. The dimensionless tip velocity is then calculated from the temporal derivative $V_{\mathrm{tip}}(t)=dh_{\mathrm{tip}}/dt$, and its evolution is compared with the microscopic-solvability-theory reference in Figure~\ref{fig:anisotropic_orient}.

\section{Conclusion}
A non-linear, Parametrised-quadratic Interface-reconstruction for the Volume-of-Fluid (PI-VOF) is presented and coupled with a geometric normal-velocity extension procedure for evolving interfaces. The formulation constructs local quadratic curves constrained by the prescribed volume fractions and interface normals, providing a higher-order geometric representation which is used to extend the interface-normal velocity into a narrow band without solving an auxiliary extension equation.
A series of benchmark problems was used to evaluate reconstruction accuracy, conservative advection, and interface evolution. Static reconstruction tests demonstrate higher-than-second-order convergence in symmetric-difference area and near-second-order convergence in curvature. Reversible-vortex advection retains approximately second-order accuracy with relative-mass conservation to numerical tolerance on refined grids. Prescribed circular expansion demonstrates approximately near-second-order convergence in both radius and geometric error, while curvature-dependent propagation establishes convergent interface evolution with the reconstructed curvature and extended velocity.

The formulation is further coupled with a ghost-fluid discretisation for Stefan problems involving interfacial heat transfer, capillarity, and attachment kinetics. The Frank circle and curvature–kinetic benchmarks demonstrate approximately second-order thermal convergence and convergent interface evolution under mesh refinement. Anisotropic dendritic solidification further demonstrates orientation-dependent growth, with tip velocities evaluated against a semi-analytical reference from microscopic solvability theory.
Overall, the results establish parametrised-quadratic reconstruction and geometric normal-velocity extension as a viable framework for conservative VOF-based interface propagation and sharp-interface simulations involving curvature-dependent and Stefan-type moving boundaries.

\bibliographystyle{ieeetr}
\bibliography{refs}

@article{schmidt1994,
title = {Computation of three dimensional dendrites with finite elements},
author   = {A Schmidt},
journal  = {Journal of computational physics},

year  = {1996},}

@article{Kim20000,
  title   = {Computation of dendritic microstructures using a level set method},
  author  = {Kim, Y. T. and Goldenfeld, Nigel and Dantzig, J. A.},
  journal = {Physical Review E},
  volume  = {62},
  number  = {2},
  pages   = {2471--2474},
  year    = {2000},
  doi     = {10.1103/PhysRevE.62.2471}
}

@article{Nagata2005,
title = {Smooth Local Interpolation of Surfaces Using Normal Vectors},
author   = {Takashi Nagata},
journal  = {Computer Aided Geometric Design},
year  = {2005},}

@article{Pilliod2004,
  title={Second-order accurate volume-of-fluid algorithms for tracking material interfaces},
  author={Pilliod Jr, JE and Puckett, EG},
  journal={Journal of Computational Physics},
  volume={199},
  number={2},
  pages={465--502},
  year={2004},
  publisher={Elsevier}
}

@article{Karagadde2024,
  author={Abhishek, G S and Karagadde, S},
  title={A hybrid marching cubes based IsoAlpha method for interface reconstruction},
  journal={Int. J. Numer. Methods Fluids},
  year={2024},

}

@article{Scardovelli2003,
  author    = {Scardovelli, R and Zaleski, S},
  title     = {Interface reconstruction with least-square fit and split Eulerian--Lagrangian advection},
  journal   = {International Journal for Numerical Methods in Fluids},
  volume    = {41},
  number    = {3},
  pages     = {251--274},
  year      = {2003},
  doi       = {10.1002/fld.428}
}

@article{Ye2023,
  author    = {Ye, Z},
  title     = {A comparative study of split advection algorithms on the moment-of-fluid (MOF) method for incompressible flow},
  journal   = {Computers \& Fluids},
  volume    = {255},
  pages     = {105815},
  year      = {2023},
  doi       = {10.1016/j.compfluid.2023.105815}
}

@article{Popinet2009,
  author={Popinet, S.},
  title={An Accurate Adaptive Solver for Surface-Tension-Driven Interfacial Flows},
  journal={J. Comput. Phys.},
  year={2009},
  volume={228},
  pages={5838--5866}
}

@article{hirt1981volume,
  title={Volume of fluid (VOF) method for the dynamics of free boundaries},
  author={Hirt, C. W. and Nichols, B. D.},
  journal={Journal of computational physics},
  volume={39},
  number={1},
  pages={201--225},
  year={1981},
  publisher={Elsevier}
}

@article{osher1988fronts,
  title={Fronts propagating with curvature-dependent speed: Algorithms based on Hamilton–Jacobi formulations},
  author={Osher, S and Sethian, J A},
  journal={Journal of Computational Physics},
  volume={79},
  number={1},
  pages={12--49},
  year={1988}
}

@article{sussman1994level,
  title={A level set approach for computing solutions to incompressible two-phase flow},
  author={Sussman, M and Smereka, P and Osher, S},
  journal={Journal of Computational Physics},
  volume={114},
  number={1},
  pages={146--159},
  year={1994}
}

@book{prosperetti2009multiphase,
  title={Computational methods for Multiphase flow},
  author={Prosperetti, A and Tryggvason, G},
  journal={Multiphase Science and Technology},
  publisher={Cambridge University Press},
  year={2009}
}

@book{tryggvason2011direct,
  title={Direct numerical simulations of gas--liquid multiphase flows},
  author={Tryggvason, G},
  year={2011},
  publisher={Cambridge University Press}
}

@article{scardovelli1999interface,
  title={Direct Numerical Simulation of Free-Surface and Interfacial Flow},
  author={Scardovelli, R and Zaleski, S},
  journal={Annual review of fluid mechanics},
  volume={31},
  number={1},
  pages={567--603},
  year={1999}
}

@article{fedkiw1999non,
  title={A non-oscillatory Eulerian approach to interfaces in multimaterial flows (the ghost fluid method)},
  author={Fedkiw, R P and Merriman, B and  Osher, S and Aslam, T },
  journal={Journal of Computational Physics},
  volume={152},
  number={2},
  year={1999},
  publisher={Elsevier}
}

@book{li2006immersed,
  title        = {The Immersed Interface Method: Numerical Solutions of PDEs Involving Interfaces and Irregular Domains},
    Author = {Li, Z and Ito, K},
  series       = {Frontiers in Applied Mathematics},
  volume       = {33},
  year         = {2006},
  publisher    = {Society for Industrial and Applied Mathematics (SIAM)},
  address      = {Philadelphia, PA},
  doi          = {10.1137/1.9780898717464},
  isbn         = {978-0-89871-609-2}
}

@article{ghost2002,
  title={A Second-Order-Accurate Symmetric Discretization of the Poisson Equation on Irregular Domains},
  author={Gibou, F and  Fedkiw, R P and  Cheng, L and  Kang, M},
  journal={Journal of Computational Physics},
  volume={176},
  number={1},
  year={2002},
  publisher={Elsevier}
}

@article{Lopez2024Isosurface,
  author       = {L{\'o}pez, J},
  title        = {Isosurface extraction for piecewise-linear reconstruction of complex interfaces on arbitrary grids},
  journal      = {Computer Methods in Applied Mechanics and Engineering},
  volume       = {425},
  pages        = {116951},
  year         = {2024},
  doi          = {10.1016/j.cma.2024.116951},
  url          = {https://doi.org/10.1016/j.cma.2024.116951}
}

@article{level-set-review-gibou-2018,
title = {A review of level-set methods and some recent applications},
journal = {Journal of Computational Physics},
volume = {353},
pages = {82-109},
year = {2018},
issn = {0021-9991},
doi = {https://doi.org/10.1016/j.jcp.2017.10.006},
url = {https://www.sciencedirect.com/science/article/pii/S0021999117307441},
author = {Frederic Gibou and Ronald Fedkiw and Stanley Osher},
}

@article{SHASHKOV-2023-mof,
title = {Moments-based interface reconstruction, remap and advection},
journal = {Journal of Computational Physics},
volume = {479},
pages = {111998},
year = {2023},
issn = {0021-9991},
doi = {https://doi.org/10.1016/j.jcp.2023.111998},
url = {https://www.sciencedirect.com/science/article/pii/S0021999123000931},
author = {Mikhail Shashkov and Eugene Kikinzon},
}

@article{quinghai-4th-track-2014,
author = {Zhang, Qinghai and Fogelson, Aaron},
title = {Fourth-Order Interface Tracking in Two Dimensions via an Improved Polygonal Area Mapping Method},
journal = {SIAM Journal on Scientific Computing},
volume = {36},
number = {5},
pages = {A2369-A2400},
year = {2014},
doi = {10.1137/140951886},

}

@book{crank1987free,
  author    = {Crank, John},
  title     = {Free and Moving Boundary Problems},
  year      = {1987},
  publisher = {Clarendon Press},
  address   = {Oxford; New York},
  language  = {English}
}

@article{lopez2013twogrid,
  author  = {L{\'o}pez, J. and G{\'o}mez, P. and Hern{\'a}ndez, J. and Faura, F.},
  title   = {A Two-Grid Adaptive Volume of Fluid Approach for Dendritic Solidification},
  journal = {Computers \& Fluids},
  volume  = {86},
  pages   = {326--342},
  year    = {2013},
  doi     = {10.1016/j.compfluid.2013.07.017},
  publisher = {Elsevier}
}

@article{lopez2010vof_JCP_solid,
  author    = {L{\'o}pez, J. and G{\'o}mez, P. and Hern{\'a}ndez, J.},
  title     = {A Volume of Fluid Approach for Crystal Growth Simulation},
  journal   = {Journal of Computational Physics},
  volume    = {229},
  number    = {19},
  pages     = {6663--6672},
  year      = {2010},
  doi       = {10.1016/j.jcp.2010.05.026},
  publisher = {Elsevier}
}

@article{Gibou2003_JSC,
  author    = {Gibou, F. and Fedkiw, R. and Caflisch, R. and Osher, S.},
  title     = {A Level Set Approach for the Numerical Simulation of Dendritic Growth},
  journal   = {Journal of Scientific Computing},
  volume    = {19},
  number    = {2},
  pages     = {183--199},
  year      = {2003},
  month     = {December},
  doi       = {10.1023/A:1025399807998},
  url       = {https://doi.org/10.1023/A:1025399807998}
}

@article{Aslam1996,
  author  = {Aslam, Tariq D. and Bdzil, John B. and Stewart, D. Scott},
  title   = {Level Set Methods Applied to Modeling Detonation Shock Dynamics},
  journal = {Journal of Computational Physics},
  volume  = {126},
  number  = {2},
  pages   = {390--409},
  year    = {1996},
  doi     = {10.1006/jcph.1996.0145}
}

@unpublished{MinnikantiEICRUnderReview,
  author = {Minnikanti, V. M. S. K. and Yan, Wenyi and Karagadde, Shyamprasad},
  title  = {An Enriched Iso-Contour Reconstruction in Volume-of-Fluid Method for Multi-Material Interfaces and Irregular Domains},
  note   = {Under review}
}

@article{Reitzle2017,
  author  = {Reitzle, M. and Kieffer-Roth, C. and Garcke, H. and Weigand, B.},
  title   = {A volume-of-fluid method for three-dimensional hexagonal solidification processes},
  journal = {Journal of Computational Physics},
  volume  = {339},
  pages   = {356--369},
  year    = {2017},
  doi     = {10.1016/j.jcp.2017.03.001}
}

@article{Fakhreddine2024,
  author  = {Fakhreddine, Ali and Alam{\'e}, Karim and Mahesh, Krishnan},
  title   = {Directionally-split volume-of-fluid technique for front propagation under curvature flow},
  journal = {International Journal for Numerical Methods in Fluids},
  volume  = {96},
  number  = {9},
  pages   = {1517--1554},
  year    = {2024},
  doi     = {10.1002/fld.5312}
}

@techreport{Kraft1988SLSQP,
  author      = {Kraft, Dieter},
  title       = {A Software Package for Sequential Quadratic Programming},
  institution = {Deutsche Forschungs- und Versuchsanstalt f{\"u}r Luft- und Raumfahrt (DFVLR)},
  number      = {DFVLR-FB 88-28},
  address     = {Oberpfaffenhofen, Germany},
  year        = {1988}
}

\end{document}